\documentclass[12pt]{article}
\usepackage{amsfonts,amsmath}
\usepackage{amssymb}

 \makeatletter \@addtoreset{equation}{section} \makeatother

 \def\theequation{\thesection.\arabic{equation}}

\newcommand{\Oo}{{{K}}}

\newcommand{\NNXY}{{\TT\htimes \HHH(y_+)}}
 \newcommand{\HHH}{{\mathbf{H}}}

\newcommand{\PH}{{  {PH_l} }}
\newcommand{\HG}{H_{M} }
\newcommand{\HGC}{{}^{\mathbb{C}}H_{M} }

\newcommand{\FFf}{\mathcal{F}}
\newcommand{\ff}{\large{\texttt{f}}}

\newcommand{\QQ}{\mathbf{Q}}

 \newcommand{\hhh}{{\nu^{-1}}}%
 \newcommand{\hhhh}{ \nu}%

 \newcommand{\be}{\begin{equation}}
\newcommand{\ee}{\end{equation}}
\newcommand{\bee}{\begin{eqnarray}}
\newcommand{\beee}{\begin{array}}
\newcommand{\eee}{\end{eqnarray}}
\newcommand{\eeee}{\end{array}}

\newcommand{\gp}{\varrho}
\newcommand{\gn}{\nu}
\newcommand{\gm}{\mu}

\newcommand{\ga}{\alpha}
\newcommand{\gb}{\beta}

\newcommand{\gs}{\sigma}

\newcommand{\M}{{\cal M}}

\newcommand{\A}{{\cal A}}

\newcommand{\B}{{\cal B}}
\newcommand{\F}{{\cal F}}
\newcommand{\E}{{\cal E}}
\newcommand{\ls}{\!\!\!\!\!\!}
\newcommand{\htimes}{\subset{\ls \times}}

\newcommand{\FF}{F}

\newcommand{\dis}{\displaystyle}

\newcommand{\gvep}{\varepsilon}
\newcommand{\gd}{\delta}

\newcommand{\go}{\omega}

\newcommand{\q}{\,,\qquad}
\newcommand{\ie}{{\it i.e.,} }

\newcommand{\Xp}{{T^{(r)}}}

\newcommand{\II}{{\mathcal R}}
\newcommand{\XR}{{L}}
\newcommand{\XRp}{{T^{(l)}{}}}

\newcommand{\nn}{\nonumber}
\newcommand{\half}{\frac{1}{2}}
\newcommand{\chalf}{\frac{1}{4}}

\newcommand{\p}{\partial}

\newcommand{\N}{{\cal N}}
\newcommand{\D}{{\cal D}}

\newcommand{\C}{{\cal C}}
\newcommand{\Q}{{\cal Q}}

\newcommand{\f}{\frac}
\newcommand{\TT}{  \mathbf{T}}
\newcommand{\G}{{\tilde \xi}}

\renewcommand{\G}{\mathcal{G}}

\renewcommand{\gp}{{P}}
\renewcommand{\gn}{{N}}
\renewcommand{\gm}{{M}}

\renewcommand{\ga}{{A}}
\renewcommand{\gb}{{B}}

\renewcommand{\gs}{{S}}

\usepackage{hyperref}

\begin{document}

\begin{flushright}
\vspace{1mm} FIAN/TD/30-09\\
 December {2009}\\
\end{flushright}

\vskip1.5cm

 \begin{center}
 {\large\bf
 Unfolding versus BRST and currents in\\ $Sp(2M)$ invariant higher-spin theory}
 \vglue 0.6  true cm

\vskip0.5cm

 O.A. Gelfond$^1$ and M.A.~Vasiliev$^2$
 \vglue 0.3  true cm

 ${}^1$Institute of System Research of Russian Academy of Sciences,\\
 Nakhimovsky prospect 36-1,
 117218,
 Moscow, Russia

 \vglue 0.3  true cm

 ${}^2$I.E.Tamm Department of Theoretical Physics, Lebedev Physical
 Institute,\\
 Leninsky prospect 53, 119991, Moscow, Russia

 \end{center}

\vskip2cm

 \begin{abstract}
The correspondence between BRST and unfolded formulations of field
equations on group manifolds and homogeneous spaces is described.
The previously introduced nonstandard BRST operator, that underlies
$Sp(2M)$ invariant higher-spin field equations, is shown to admit a
natural oscillator-like realization. The coordinate independent form
of conserved currents in the $Sp(2M)$ invariant higher-spin theory
is derived from the BRST formulation on $Sp(2M)$ extended by the
Heisenberg group.
 \end{abstract}
\newpage
\tableofcontents
\newpage

\newpage
\section{Introduction}

In \cite{gelbrst} it was shown that the unfolded
formulation of \cite{BHS} of $Sp(2M)$ invariant higher-spin (HS)
theories  \cite{F,BLS1} (see also
\cite{Mar,cur,IB,DV,PST,tens2,BPST,BBAST,EL,EI,33,gelcur}) can be
equivalently formulated as a BRST closure condition for a certain
nonstandard BRST operator $Q$ on the semidirect product $SpH(2M )
=Sp(2M ) {\htimes} \HG$ where $\HG$ is the Heisenberg group. The aim
of this paper is to clarify the invariant origin of this relation
established in \cite{gelbrst} in a particular coordinate system.

We start with the discussion of  the general relation between the
BRST and unfolded formulations, explaining in a
coordinate-independent way that the two approaches are essentially
equivalent once the BRST formulation is given in terms of Lie vector
fields on a group manifold. The relation via identification of the
ghost fields of the BRST formulation with the differential forms of
the unfolded formulation requires however a nontrivial relation
between  space derivatives in the two formulations.

The proposed approach is applicable to dynamical systems formulated
in nontrivial geometries of group manifolds and homogeneous spaces
as well to their deformations. In particular, it is well suited for
the extension of the BRST approaches to HS systems studied in
\cite{Barnich:2004cr,CLW} in Cartesian coordinates in flat space or
via embedding $AdS$ geometry into a higher dimensional flat space as
in \cite{BGr} to any coordinates and/or more complicated geometries.
The same time, the equivalence between the unfolded formulation and
appropriately interpreted BRST formulation in the proposed setup is
by construction. Among other things, this explains that the obvious
parallels between the formalisms and conclusions obtained within the
BRST approach (see e.g. \cite{CLW} for a recent work) with those
well established in the unfolded dynamics \cite{Ann, act} (and
references therein) are in no way accidental and/or surprising.

In particular, the equivalence of the two approaches explains that
BRST cohomology may describe both nontrivial deformations of field
equations and nontrivial closed differential forms in space-time,
that can be constructed from the dynamical fields of the unfolded
formulation, \ie conserved currents. The latter relation is  applied
in this paper to the further study of the BRST formulation of
$Sp(2M)$ invariant HS gauge theory of \cite{gelbrst}. Firstly, we
show that the complicated nonstandard BRST operator found in
\cite{gelbrst} has simple origin in the oscillator realization of
the symplectic algebra. Secondly, we show that the conserved charges
proposed in \cite{gelcur} correspond to certain BRST cohomology,
which observation provides their coordinate-independent realization.

The rest of the paper is organized as follows. In Section
\ref{Cartan forms} we explain the general relation between the
unfolded and BRST formulations. To make the paper selfcontained, we
recollect in Section \ref{Prel} basic formulae of \cite{gelbrst} on
the $Sp(2M)$ invariant HS theories. The new oscillator-like
realization of the nonstandard BRST operator of \cite{gelbrst} is
introduced in Section \ref{Oscillator}. The construction of closed
forms from solutions of dynamical equations is presented in Section
\ref{BRST clo} first for the general case in Subsections \ref{genc},
\ref{Parameters} and then for the $SP(2M)$ geometry in Subsections
\ref{cloM}, \ref{BRST clo3M}. The latter results are applied in
Section \ref{Bilinear} to the coordinate-independent construction of
conserved currents bilinear in HS fields, that reproduces the
previously known results in particular coordinates. Possible applications
to unfolded equations are discussed in Section \ref{conc}.

\section{BRST operators and unfolded equations}
\label{Cartan forms}

Consider a Lie group  $G $ and its  Lie algebra $\mathfrak{g}$. Let
$R_\alpha$ ($\alpha=1,\ldots dim\, G$) be right Lie vector fields on
$G$ that satisfy \be\label{comrelG} [R_\alpha\,,R_\beta
]=f_\alpha{}_\beta{}^\gamma R_\gamma\,, \ee where
$f_\alpha{}_\beta{}^\gamma$ are structure constants of
$\mathfrak{g}$. \,\, Let  $T_\alpha$ form a basis of some
representation $T$ of   $\mathfrak{g}$ \be\nn
[T_{\alpha}\,,T_{\beta} ]=f_{{\alpha}{\beta}}{}^{\gamma}
T_{\gamma}\q [R_\beta \,,T_{\alpha} ]=0\,. \ee Provided that  the
ghosts $c^{a}$ and $b_{a}$ obey the relations \be\nn [c^\alpha\,,
R_\beta]=0\q [b_\alpha\,, R_\beta]=0\q [c^\alpha\,, T_\beta]=0\q
[b_\alpha\,, T_\beta]=0 \ee \be\label{anticomrel}
\{c^{\alpha}\,,b_{\beta}\}=\delta^{\alpha}_{\beta}\q
\{c^{\alpha}\,,c^{\beta}\}=0\q \{b_{\alpha}\,,b_{\beta}\}=0\,, \ee
the BRST operator \be \label{BRST} Q= c^\alpha (R_\alpha + T_\alpha)
- \half c^{\alpha} c^{\beta} b_{\gamma}
f_{{\alpha}{\beta}}{}^{\gamma} \ee is nilpotent \be\nn
 Q^2  =0\,.
\ee

That the equation \be \label{BRSTMC} \{Q\,,c^\gamma\}=  - \half
c^{\alpha} c^{\beta}   f_{{\alpha}{\beta}}{}^{\gamma}\q
 \ee
has the   Maurer-Cartan form  upon identification of $Q$ with $d$
suggests that, being  dual to the Lie vector fields $R_\alpha$ on
$G$, the ghosts $c^\alpha$ should be identified with the Cartan
forms on $G$. However, the naive identification fails because it is
assumed that $[R_\alpha\,, c^\beta]=0$ while the Lie vector fields
$R_\alpha=R_\alpha{}^a(x) \f{\p}{\p x^a}$ do not commute to the
$x$-dependent Cartan forms ($x^a$ are coordinates on $G$).

To proceed, it is necessary  to redefine the notion of the vector
fields appropriately. Let \bee \label{cartan1} R_\alpha=
R_\alpha{}^a (x) p_a\q \\ \label{cartan11} {[p_a\,,f(x) ]= \f{\p}{\p
x^a}}f(x)\q[p_a, p_b]=0\,.
  \eee
The equation (\ref{comrelG}) amounts to the standard Lie conditions
\bee \label{comrelGp} R_\alpha{}^b(x)\,\f{\p}{\p x^b}\,R_\beta{}^a
(x) -R_\beta{}^b(x) \,\f{\p}{\p x^b}\,R_\alpha{}^a(x)\, \,
=f_\alpha{}_\beta{}^\gamma R_\gamma{}^a(x)\, \,. \eee The  Cartan
forms
 \bee \label{cartan2}
\go^\alpha=R^{-1}{}_a{}^\alpha (x)dx^a\q \eee where the
differentials $dx^a$ satisfy $\{dx^a, dx^b\}=0$, $ [dx^a, f(x)]=0$,
obey the Maurer-Cartan equation \be \label{MC} d\go^\alpha = -
\half \go^{\beta} \go^{\gamma}   f_{{\beta}{\gamma}}{}^{\alpha} \,.
\ee (We systematically skip the wedge symbol throughout this paper.)
The key point is to set \bee
 \label{covd} p_n&=&
   \f{\p}{\p x^n} - c^\alpha\,b_\beta\, R_\alpha^m
    \f{\p}{\p x^n} (R^{-1}{}_m{}^\beta).
  \eee
An elementary computation shows that $p_n$ satisfies the
  relations (\ref{cartan11}). In addition, we can identify $\go^\alpha$ and
  $c^\alpha$
\be\nn \go^\alpha =c^\alpha \ee
  since
 \bee \nn [p_n\,,\go^\alpha]=
 \f{\p}{\p x^n}
(R^{-1}{}_m{}^\alpha (x)) dx^m
 -
c^\beta\, R_\beta^m \f{\p}{\p x^n} (R^{-1}{}_m{}^\alpha (x))
\Big|_{c^\gamma=\go^\gamma} =0. \eee

With this identification  we obtain from (\ref{BRST}) along with
(\ref{covd}) and  (\ref{cartan2}) \bee \label{BRSTNEW} Q=\go^\alpha
(R_\alpha + T_\alpha) - \half \go^{\alpha} \go^{\beta} b_{\gamma}
f_{{\alpha}{\beta}}{}^{\gamma}=
 \go^\alpha (R_\alpha^n \f{\p}{\p x^n}+ T_\alpha)= D\,,
\eee where $D$ is the  covariant derivative in the representation
$T$ of $G$ \be\nn
 D= d+ \go^\alpha  T_\alpha\q d=dx^n \f{\p}{\p x^n}\,.
\ee In these terms, the condition
 $Q^2=0$ is equivalent to
 \be\nn D^2=0\,,
 \ee
 which, in turn, is
the consequence of the Maurer-Cartan equation (\ref{MC}).

 Thus, the BRST closure condition
 \be\label{QC} Q \phi
=0 \ee is equivalent to the unfolded equation \be\label{UN}
 D\phi =0\,.
 \ee
Hence, $Q$ cohomology is equivalent to the $D$ cohomology. In
particular, in the case where the representation $T$ is trivial, $Q$
cohomology is equivalent to De Rham cohomology
 \bee\label{Q=d}\ls\ls
 Q  \,F(x,\,c)  =G(x,\,c)\,\,
 \Longleftrightarrow\,\,
 \,
d F(x,\,\go)   =G (x,\,\go). \eee This simple property will be used
in this paper to derive conserved HS currents from the appropriate
$Q$-cohomology.

The equivalence between the BRST approach and unfolded approach
shown in this paper contains two essential elements.

One is that the BRST operator should contain Lie vector fields of a
chosen group $G$ which form a frame of the tangent space of
$G$. As a result, the full set of derivatives on $G$ reappears in
the unfolded formulation via the exterior differential $d$.

Another one is the relation (\ref{covd}) that tells us that, for the
identification of the BRST operator $Q$ with the exterior
differential, the operator $p$ in $Q$ should be interpreted as a
covariant derivative (\ref{covd}) that acts on the space of ghosts
(forms). Note that, in accordance with the second relation in
(\ref{cartan11}),  $p_n$ (\ref{covd}) is flat. Indeed, the $Gl_{dim
\,G}$ connection in (\ref{covd}) has the standard pure gauge form
with the Lie matrix $R_n{}^\alpha(x)$ as the gauge function.

To arrive along these lines at interesting field equations  one has
to consider appropriate representations $T$ and/or further
nonstandard modifications of the BRST operator $Q$. The examples of
this construction will be considered in Section \ref{Oscillator},
 where it will be
shown in particular how the $Sp(8)$ unfolded equations of \cite{BHS}
result from the nonstandard BRST operator.

Another interesting application is to the BRST reformulation of
nonlinear HS theories that may involve higher differential
forms as dynamical variables. To this end the space of ghosts
$c^\alpha$ should be extended to a larger set $C^\ga$ that includes
objects of different non-negative degrees $p^\ga=0,1,\ldots $.
Correspondingly, the space of ghosts $b_\alpha$ extends to $B_\ga$
of degrees $-p_\ga$. The graded commutation relations are \be
[B_\ga\,,C^\gb]_\pm = \delta_\ga^\gb\q [B_\ga\,,B_\gb]_\pm =0\q
[C^\ga\,,C^\gb]_\pm =0\,, \ee where \be [a\,,b]_\pm
=ab-(-1)^{p(a)p(b)} ba\,. \ee The idea is to extend the BRST
operator as follows \be Q_\mathfrak{g} \to Q_\mathfrak{g}^\prime =
Q_\mathfrak{g} + \Q \ee where $Q_\mathfrak{g}$ is the canonical BRST
operator built from the vector fields of some group Lie $G$ while
$\Q$ is some other BRST operator built from the ghosts $C$ and $B$,
that is also nilpotent $\Q^2=0$. The field equations require the
equivalence of the action of $Q_\mathfrak{g}$ and $\Q$ on every
dynamical variable $W=W(C,B)$ \be Q_\mathfrak{g} (W) +\Q (W) =0\,.
\ee Upon the field redefinition explained above, this amounts to the
unfolded equations \be \label{un} {dW +\Q(W) =0\,,} \ee in which
form unfolded equations, introduced  originally in \cite{Ann} in the
study of HS gauge theory, were discussed more recently in \cite{act}
(see also \cite{solv,33} for more detail and references). Let us
note that in this setup $\Q$ cohomology describes nontrivial
deformation of the nonlinear equations (\ref{un}).

A remarkable feature of unfolded dynamics is that it is insensitive
to the dimension of space-time where the fields are defined. This
property is nicely illustrated within its BRST version discussed in
this paper applied to the coset space construction.

Let $H$ be a subgroup of $G$. The space of functions on $G/H$
identifies with the space of solutions of the equations \be
\label{rinv} R_{a} \FF(G) =0\,, \ee where $R_{a}$ is a subset of
right vectors field of $H$. The algebra Lie $\mathfrak{g}$ acts on
solutions of (\ref{rinv}) by the left vector fields $L_\alpha $. The
equation (\ref{rinv}) results from the restriction to the sector of
$c$--independent $\FF(G)$ of the condition \be \label{QQ}
Q_\mathfrak{h} \FF(G,c) =0\,, \ee where $Q_\mathfrak{h}$ is the
canonical BRST operator of $H$ and $b_{a}$ is realized as $
\f{\p}{\p c^{a}}\,. $

The extension of (\ref{rinv})  to the induced module construction
is \be \label{ind} (R_{a}+T_{a}) \FF(G) =0\,, \ee where $\FF(G)$ is
valued in some $H$-module $V$ and $T_{a}$ provide a representation
of the Lie algebra $\mathfrak{h}$ of $H$ on $V$.
 In what
follows we will be interested in the particular case
 with a  one dimensional $H$--module $V$. In this case,
$\FF(G)$ is still valued in $\mathbb{R}$ or $\mathbb{C}$ and $T_{a}$
is given by some constants associated to central elements of the
grade zero  part of   $\mathfrak{h}$. The equation (\ref{ind})
results from the restriction to the sector of $c$--independent
$\FF(G)$ of the condition (\ref{QQ}). The BRST  extension of
the equation (\ref{ind}) to $\FF(G,c)$  is
conveniently interpreted in terms of the Fock module generated from the
vacuum that satisfies $b_{a}|0\rangle=0$.

In the unfolded dynamics approach, the phenomenon illustrated by the
example  of  a coset manifold, that a theory in a
smaller space $G/H$ can be described as that in a larger space
$G$, extends to less symmetric situations\footnote{Strictly
speaking this is true for the so-called universal unfolded systems
\cite{solv} which case is however general enough to cover all known
examples of unfolded equations.} where the symmetry $G$ is broken or
deformed. As a result, the concept of space-time dimension turns out
to be dynamical in unfolded dynamics.

To apply this approach to the analysis of the $Sp(2M)$ invariant HS
field equations let us first recall main ingredients of the
formalism of \cite{gelbrst}.

 \section{Preliminaries}
\label{Prel}
\subsection
{Heisenberg extension of  symplectic group} \label{Heisenberg
extension}

The group $Sp(2M|\mathbb{R})$ is constituted by real matrices \bee
\label{sp2M} \beee{c r } G= & \left( \beee{r r}
a^\ga{}_\gb&b^{\ga\gm}\\c_{C\gb}&d_C{}^\gm  \eeee \right)
 \eeee
{} \eee with $M\times M$ blocks
$a^\ga{}_\gb\,,\,b^{\ga\gb}\,,\,c_{\ga\gb}\,,\,d_\ga{}^\gb$ that
satisfy the relations \bee\label{sprel1}\!\! a{}^{\ga}{}_{C
}b{}^{{D}\,C}- a{}^{D}{}_C b{}^\ga{}^C=0,\quad
 a{}^{\ga}{}_{C } d{}_B{}^{\,C}-b{}^\ga{}^C c{}_{ B \,C}=\gd^{\ga}{}_{ B}\,,
\quad   c{}_{B C} d{}_A{}^C- c{}_{A C} d{}_B{}^C   =0\,\, \eee
 equivalent to the invariance condition $ARA^t = R$
for the symplectic form $R= \left( \begin{array}{cc}
                   0    &   I^{\ga}{}_{\gb}   \\
                   -I_{C}{}^{D}   &   0
            \end{array}
    \right)\,,
$ where $I$ is the  unit $M\times M$ matrix and $A^t$ is the
transposed matrix.

Any $g\in Sp(2M|\mathbb{R})$  with nondegenerate
 $d$ (\ref{sp2M}) can be represented in the form
\bee \label{spprod} \beee{c c r r } \left( \beee{r r}
a^{A}{}_{B}&b^{{A}{C}}\\c_{{D}{B}}&d_{D}{}^{C}  \eeee \right)   =&
 \left( \beee{r r} \gd^{A}{}_{E}&  X{}^{A}{}^{F}
\\  0&\gd_{D}{}^{F}  \eeee \right)
& \left( \beee{r r} \A{}^{E}{}_{G}&\quad 0
\\\quad 0&\D_{F}{}^{H}  \eeee \right)
& \left( \beee{r r} \gd^{G}{}_{B}&0
\\ \C_{ H}{}_{B}&\gd_{H}{}^{C}  \eeee \right).
\eeee & \eee This gives \bee \nn \beee{  r r } \left( \beee{r r}
a^{A}{}_{B}&b^{{A}{C}}\\c_{{D}{B}}&d_{D}{}^{C}  \eeee \right)   =&
 \left( \beee{r r} \A^A{}_{B}
 +X^{\ga {F}}\D_{F}{}^{G}\C_{{G}{B}}
   &\quad X^{\ga {F}}\D_{F}{}^{C}\\
    \D_{D}{}^{G}  \C_{{G}{B}}
   & \D_{D}{}^{C}   \eeee \right)\,,
\eeee \eee where \bee\label{coorsp}\A{}^{A}{}_\gb
=(d^{-1}){}_\gb{}^{A}\,,\quad X^{\ga\,\gm} =b{}^\ga{}^{C}
\A{}^\gm{}_{C} , \quad \C_{{B}\,{A}}=c_{{C}{B}} \A{}^{C}{}_{A} \eee
can be chosen as local coordinates on $Sp(2M|\mathbb{R} )$. Note
that $X^{\gb\,\ga}=X^{\ga\,\gb}$ and $\C_{\gb\,\ga}=\C_{\ga\,\gb}$
by virtue of the identities \bee\nn\quad
 - c{}_{\,{B}}{}_{{A}}d^{-1}{}_{{C} }{}^{B} +
c{}_{\,{B}}{}_{{C} }  d^{-1}{}_{A}{}^{B}=0\,,\quad  -b{}^\ga{}^{B}
d^{-1}{}_{B}{}^{C}+b{}^{{C}}{}^{\,{B}}d^{-1}{}_{{B} }{}^{\ga}=0,
\eee which follow from (\ref{sprel1}). Note that
$\D{}_{A}{}^{B}=d{}_{A}{}^{B}\,. $

 $Sp(2M|\mathbb{R})$ contains the following important subgroups.
 The Abelian subgroup of translations  $\TT$ consists of  the elements
\bee t(X)= \left( \begin{array}{cc}
                   I    &   X   \\
                   0    &   I
            \end{array}
    \right)\,
\label{4} \eee with various  $X^{\ga\gb}=X^{\gb\ga}$. The product
law in $\TT$ is $ t(X)t(Y) = t(X+Y)$.
\newcommand{\SSS}{   \mathrm{S}}

Analogously, the Abelian subgroup $\SSS$ of special conformal
transformations is constituted by the matrices (\ref{sp2M}) with
$a=d=I$, $b=0$. The subgroup $GL(M)$ of generalized Lorentz
transformations $SL(M)$ and dilatations    consists of the matrices
(\ref{sp2M}) with $b=c=0$ and $a^{\gb}{}_{{C}} d_\ga{}^{C}
=\delta_\ga^\gb\,.$

The lower $P_l(\mathbb{R}) $ and upper $P_u(\mathbb{R}) $ maximal
parabolic subgroups  of $Sp(2M|\mathbb{R})$ are
 \be P_l \ni
p= \left( \begin{array}{cc}
                   a    &   0  \\
                   c   &   d
            \end{array}
    \right)\,\q P_u \ni
p= \left( \begin{array}{cc}
                   a    &   b  \\
                   0   &   d
            \end{array}
    \right)\,.
\label{p} \ee

 Let $\HG =\mathbb{R}^M\times \mathbb{R}^M \times\mathbb{R}^1$
be the $(2M+1)-$dimensional Heisenberg group constituted by
\bee\label{Heis} \FFf=\{\ff \,,\,u\}\,,\qquad \ff=y^\ga\,,\,w_\gb\,,
\qquad \ga,\,\,\gb=1,...,M \eee with the product law
$$\FFf_1\circ\,\FFf_2=\{\ff_1{}+\ff_2{} \,,\, \,u_1+u_2-
( \ff_1\,, \, \ff_2)\}\,,$$ where  $(\,,\,)$ is the symplectic form
\bee\label{simfor} ( \ff_1\,, \,\ff_2)
=y_1{}^\ga\,\,w_2{}{}_\ga-y_2{}^\ga\,\,w_1{}_\ga\,=-( \ff_2\,,
\,\ff_1)\,, \qquad \ga = 1,\ldots, M . \eee

  $Sp(2M|\mathbb{R})$ \Big($Sp(2M|\mathbb{C})$\Big) acts canonically on  $\HG$
\Big($\HGC$\Big) which is the manifestation of the standard fact
that $Sp(2M)$ possesses the oscillator realization (see e.g.
\cite{BG}). This makes it possible to introduce the group $SpH(2M )
=Sp(2M ) \htimes \HG$
   \bee\label{Sph}
SpH(2M )\,:\,\G=\{G \,,\,\FFf  \}\,,\qquad 
G\in Sp(2M),\quad \FFf=\{\ff \,,\,u\}\in  \HG \eee with the product
law \be\nn \G_1\circ \G_2=\{ G_1 G_2 \,,\,\,\, \ff_1\,+\,G_1 \ff_2\,
\,,\,\,\,u_1\,+\,u_2\,-\,(\ff_1\,,G_1 \ff_2)\}\,, \ee where
$(\,,\,)$ is the symplectic form (\ref{simfor}) and \bee\nn \beee{c
r l r c } G   \ff= &\ls \left(\! \beee{r r}
                         a^\ga{}_\gb&b^{\ga\gm}\\c_{{C}\gb}&d_{C}{}^\gm
                          \eeee \!\right)
&
 \ls\left(\!\! \beee{r  }
y{}^\gb \\w_{\gm}   \eeee\!\! \right)= &\ls\, \left( \!\beee{r }
              a^\ga{}_\gb y{}^\gb+ b^{\ga\gm}w_{\gm}
              \\
               c_{CB}y{}^\gb+d_{C}{}^\gm w_{\gm}
          \eeee \!\!\right),
&\!\!
  G \in Sp(2M),\,\,\,  \ff =(y^\ga,w_\gb)\in\HG.
\eeee \eee

Analogously,
 a rank $ r$ Heisenberg extension   $ SpH_r(2M |\mathbb{A})$ is
 introduced  for any field $\mathbb{A}$ and $r\in \mathbb{N}$ as
   \bee\label{semi}
SpH_r(2M|\mathbb{A} )=Sp(2M|\mathbb{A}) \htimes \underbrace{
\HG(\mathbb{A})\times\dots\times \HG(\mathbb{A})}_r
 \eee
 with coordinates
 \bee\label{coorHsr}
y_j{}^\ga\q\,w_j{}_\gb\q u_j{}\q j=1,\dots,r.
 \eee
When it does not lead to misunderstandings, we will use  shorthand
 notation like  $SpH$  instead of $SpH(2M|\mathbb{R})$ etc.
 Note, that $SpH_1(2M )\equiv SpH(2M )$ and
 $SpH_0(2M )\equiv Sp(2M )$.

The  lower quasiparabolic  subgroup $\PH
(2M|\mathbb{R})=P_l{}\htimes H^-
 \subset SpH(2M|\mathbb{R})$ consists of the elements
 \bee \nn\label{parab}
\beee{c r c } \PH (2M|\mathbb{R})  =\left\{\rule{0pt}{18pt}\right.&
\left( \beee{c c} p^\ga{}_\gb&0\,\\  p_{C}{}_\gb&p_{C}{}^D  \eeee
\right), &0\,,\,p_\ga\,,\,p\left.\rule{0pt}{18pt}\right\}\,. \eeee
\eee Possible local coordinates on   $SpH/ \PH $ are \,\, \be
X^{\ga\gb}\,\, ,\quad Y^\ga=y^\ga- w_{\gb}X^{\ga\gb} \label{XY}.\ee
Analogously we define $SpH_r/ \PH_r$ with local coordinates $
X^{\ga\gb}\,$ and\,\, $Y_1{}^\ga,\dots,Y_r{}^\ga$\,. In the case of
$r=0$, this gives the Lagrangian Grassmannian with $X^{AB}$ being
 local coordinates of its big cell $\M_M$. Indeed, from Eq.~(\ref{spprod})
it follows that any element (\ref{sp2M}) of $Sp(2M|\mathbb{R})$ with
$det |d^\ga{}_\gb | \neq 0$, which condition singles out the big
cell of the Lagrangian Grassmannian, belongs to some equivalence
class associated to a point of  $\M_M$.

\subsection{Vector fields}
\label{Vector fields}

Any Lie group $G$ possesses two mutually commuting sets of  left
 and right Lie
vector fields $\XR_{\beta}$ and $R _{\alpha}$
(${\alpha}\,,\,\,{\beta}=1,2,\ldots, \dim G$), where indices
${\alpha}\,,{\beta}\,,\ldots$ enumerate a basis of $G$, each forming
the Lie algebra $\mathfrak{g}$ of $G$ \be\label{comleftright}
[R_{\alpha} \,,R_{\beta}  ]=f_{{\alpha}{\beta}}{}^{\eta} R_{\eta} \q
[\XR_{\alpha} \,,\XR_{\beta} ]=f_{{\alpha}{\beta}}{}^{\eta}
\XR_{\eta} \q [R_{\alpha} \,,\XR_{\beta}  ]=0\,. \ee

The straightforward calculation of $Sp(2M)$ right vector fields in
the coordinates (\ref{coorsp}), (\ref{coorHsr})
gives 
  \bee \label{vect_fields sp}
R_{\ga\,\gb}&=&
 -2 \A{}^{ {E} }{}_{\ga}\A{}^{{D} }{}_{\gb}\f{\p}{\p X^{{D}{E}}}
+   2\A{}^{{E}}{}_{(\gb}\C_{ \ga){D} }\f{\p}{\p \A{}^{{E}}{}_{{D}}}
+2\C_{\ga{D}}\C_{\gb{E}}\f{\p}{\p \C_{{D} {E}}} ,\rule{18pt}{0pt}\\
\nn R^{A\,B}&=& 2 \f{\p}{\p \C_{A\,B}},\\ \nn
R_A{}^{B}&=&-2\C_{A{C}}\f{\p}{\p \C_{B{C}}}
               -\A{}^{{C}}{}_{A}\f{\p}{\p \A{}^{{C}}{}_{B}}\,.
\eee
  $SpH_r(2M)$   right
vector fields contain in addition the Heisenberg  vector fields
\bee\label{rightgzr} R_{j}{}^{{C}}&=&\D{}_\gm{}^{C}\Big(
X{}^\gm{}^{A}   \f{\p}{\p y_{j}{}^{{A}}} +\f{\p}{\p
w_{j}{}_{\gm}}+\Big(- y_{j}{}^{\gm}+
 w_{j}{}_{{A}}  X{}^\gm{}^{A}
\Big)\f{\p}{\p u_{j}{}}\Big),
\\ \nn
 R_{j}{}_{\ga}&=&
- \A{}^\gm{}_\ga\Big( \f{\p}{\p y_{j}{}^{\gm}}   +
 w_{j}{}_{\gm}  \f{\p}{\p u_{j}{}}\Big)
-\C{}_\ga{}_\gn R_{j}{}^{\gn},
\\ \nn
R_{j}{}&=&2\f{\p}{\p u_{j}{}}, \eee where $j=1,...,r.$ Note, that
for $r=1$ the index $j$ will be omitted.

In the same local coordinates, the left Heisenberg vector fields are
\bee \label{leftgzr} \XR_{j}{}_{\ga}&=&\f{\p}{\p y_{j}{}^{\ga}} -
w_{j}{}_{\ga}\f{\p}{\p u_{j}{}}
\q
\XR_{j}{}^{\ga}=\f{\p}{\p w_{j}{}_{\ga}}+   y_{j}{}^{\ga}\f{\p}{\p
u_{j}{}}
\q
\XR_{j}{}=2\f{\p}{\p u_{j}{}}.
  \eee

 Nonzero commutation relations of (right) vector fields of $SpH_r$ are
\bee\nn [R^\ga{}_\gb \,,R^{C}{}_{E}] &=& \delta_\gb^{C} R^\ga{}_{E}
-\delta_{E}^\ga R^{C}{}_\gb\,,
\\\nn
[R^\ga{}_\gb \,,R^{{C}{E}}] &=& \delta_\gb^{C} R^{\ga{E}} +
\delta_\gb^{E} R^{\ga{C}}\,,
\\   \label{comrelsp1}
[R^\ga{}_\gb \,,R_{{C}{E}}] &=& -\delta^\ga_{C} R_{\gb{E}} -
\delta^\ga_{E} R_{\gb{C}}\,,
\\  \nn
[R_{\ga\gb}\,,R^{{C}{E}} ] &=& \delta_\ga^{C} R^{E}{}_\gb+
\delta_\gb^{C} R^{E}{}_\ga +\delta_\ga^{E} R^{C}{}_\gb+
\delta_\gb^{E} R^{C}{}_\ga \, \eee and \bee   \label{crsph1}
[R^\ga{}_\gb \,,R_j{}^{{C}}] &=& \delta_\gb^{C}
R_j{}^{\ga}\q\qquad\,\,\quad [R^\ga{}_\gb \,,R_j{}_{{C}}] =
-\delta^\ga_{C} R_j{}_{\gb}\,,
\\ \nn
[R_{\ga\gb}\,,R_j{}^{{C}} ] &=&
 \delta_\gb^{C} R_j{}_\ga{}+\delta_\ga^{C} R_j{}_\gb{}\q
[R^{\ga\gb}\,,R_j{}_{{C}} ] =
 -\delta^\gb_{C} R_j{}{}^\ga{}-\delta^\gb_{C} R_j{}{}^\gb{}\,,
\\  \nn
[R_k{}{}_\ga\,,R_j{}{}^\gb ] &=& \delta_{k j}  \delta_\ga^\gb R_j{}
\,.\eee

In the case of $r=2$ it is convenient to introduce
 \bee   \label{sphpm}
 \beee{l l l}   y_\pm{}^A=y_1{}^A\pm y_2{}^A,& w_\pm{}_A=(w_1{}_A\pm w_2{}_A),
 & u_\pm=(u_1\pm u_2) ,
 \\   R_\pm{}^A=\half(R_1{}^A\pm R_2{}^A),&
 R_\pm{}_A=\half(R_1{}_A\pm R_2{}_A)\,,& R_\pm{} =\half(R_1{} \pm R_2 )\,.\rule{0pt}{20pt}
 \eeee
 \eee
Eqs.(\ref{rightgzr}) acquire the form
  \bee\label{gz-F}
R_{-}{}^{{C}}&=&\D{}_D{}^{C}\Big(\f{\p}{\p w_{-}{}_{D}} +
X{}^D{}^{A}
 \f{\p}{\p y_-{}^A}
-\half Y_{+}{}^{D} \f{\p}{\p u_{-}{}}
 -\half Y_-{}^{D} \f{\p}{\p u_{+}{}}
\Big),
\\ \nn
 R_{-}{}_{\ga}&=&
- \A{}^D{}_\ga \Big( \f{\p}{\p y_-{}^{D}}
 +\half w_{-}{}_{D} \f{\p}{\p u_{+}{}}+
\half w_{+}{}_{D} \f{\p}{\p u_{-}{}}\Big)
  -\C{}_\ga{}_D R_{-}{}^{D},
\\ \nn
R_{-}{}&=&2\f{\p}{\p u_{-}{}}, \eee \bee\label{gz+F}
R_{+}{}^{{C}}&=&\D{}_D{}^{C}\Big(\f{\p}{\p w_{+}{}_{D}} +
X{}^D{}^{A}
  \f{\p}{\p y_{+}{}^{{A}}}
-\half Y_{+}{}^{D} \f{\p}{\p u_{+}{}} -\half Y_{-}{}^{D} \f{\p}{\p
u_{-}{}}\Big) ,
\\ \nn
 R_{+}{}_{\ga}&=&
- \A{}^D{}_\ga\Big( \f{\p}{\p y_{+}{}^{D}} +\half  w_{+}{}_{D}
\f{\p}{\p u_{+}{}} +\half   w_{-}{}_{D}  \f{\p}{\p u_{-}{}}\Big)
-\C{}_\ga{}_D R_{+}{}^D,
\\ \nn
R_{+}{}&=&2\f{\p}{\p u_{+}{}}, \eee
 where
$Y_{\pm}{}^{D}=y_{\pm}{}^{D}-
 w_{\pm}{}_{{A}}  X{}^D{}^{A}$.
 Note that
\bee\nn [R_\pm{}{}_\ga\,,R_\pm{}{}^\gb ] = \half  \delta_\ga^\gb
R_+{} \q [R_\pm{}{}_\ga\,,R_\mp{}{}^\gb ] = \half  \delta_\ga^\gb
R_-{} \,.\eee

The nonzero
anticommutation relations of the ghosts
of $\mathfrak{sph}_2$ are \bee
\label{antisph} \{c^{\ga\gb} \,,b_{{C}{E}}\} =\half( \delta^\ga_{E}
\delta^\gb_{C}+ \delta^\gb_{E} \delta^\ga_{C} )\,, \quad
\{c_{\ga\gb} \,,b^{{C}{E}}\} =\half( \delta_\ga^{E} \delta_\gb^{C}+
\delta_\gb^{E} \delta_\ga^{C} )\,,
\\ \nn\{c^\ga{}_\gb
\,,b^{C}{}_{E}\} = \delta^\ga_{E} \delta_\gb^{C}\,,\quad \{c^{\ga}
\,,b_{{C}}\} = \delta^\ga_\gb ,\quad \{c_i{}_{\ga} \,,b_j{}^{{C}}\}
= \delta_j^i \delta_\ga^\gb ,\quad \nn \{c \,,b\} = 1. \eee
Using conventions (\ref{sphpm}) along with
 \bee\label{sphpmcb}
  c_\pm{}_A=c_1{}_A\pm c_2{}_A\q   c_\pm{}^A=c_1{}^A\pm c_2{}^A\q c_\pm=c_1\pm c_2\q\\ \nn
  b_\pm{}^A=\half(b_1{}^A\pm b_2{}^A)\,\q  b_\pm{}_A=\half(b_1{}_A\pm b_2{}_A)\,\q b_\pm=\half(b_1\pm b_2)\,,
 \eee
one can see that the canonical BRST operator of $ SpH_{2}$\, is
\bee\label{QSpH2}  {{Q}_{SpH_2}}&=&
 c^{\ga\gb}R_{\ga\gb}
+c^\ga{}_\gb R^\gb{}_\ga   + c_{\ga\gb} R^{\ga\gb}
\\ \nn &&
+c^\ga{}_\gb c^\gb{}_{C} b^{C}_\ga -4 c^{\ga\gb} c_{\gb{C}}
b^{C}{}_\ga +2 c^\ga{}_\gb c^{\gb{C}}b_{\ga{C}}   -2c^\ga{}_\gb
c_{\ga{C}}b^{\gb{C}} \phantom{\half}
\\ \nn && +c_+ R_+{}  + c_+{}_{A} R_+{}^{A} + c_+{}^{A} R_+{}_{A}
  +c_- R_-{}     + c_-{}_{A} R_-{}^{A} + c_-{}^{A} R_-{}_{A}
\\ \nn && -c_\ga{}^\gb c_+{}_\gb b_+{}^\ga\,
+2c_\ga{}_\gb c_+{}^\gb b_+{}^\ga\, +c_\ga{}^\gb c_+{}^\ga b_+{}_\gb
-2c^\ga{}^\gb c_+{}_\ga b_+{}_\gb\, + \half c_+{}_\ga{}  c_+{}^\ga
b_+{}
\\ \nn &&-c_\ga{}^\gb c_-{}_\gb b_-{}^\ga\,
+2c_\ga{}_\gb c_-{}^\gb b_-{}^\ga\, +c_\ga{}^\gb c_-{}^\ga b_-{}_\gb
-2c^\ga{}^\gb c_-{}_\ga b_-{}_\gb\, + \half c_-{}_\ga{}  c_-{}^\ga
b_+{}\\ \nn &&+ \half c_-{}_\ga{}  c_+{}^\ga b_-{} + \half
c_+{}_\ga{} c_-{}^\ga b_-{} \,.  \eee

\subsection{Nonstandard   BRST\, operator}
\label{Nonstandard}

Let some operators $P_\alpha$ form a ``closed algebra"
 \bee\label{brack} [P_{\alpha},
P_\beta]=\phi_{\alpha\beta}^\gamma (R) P_\gamma \eee where both
$P_\alpha (R)$ and ``structure functions" $\phi_{\alpha\beta}^\gamma
(R)$ belong to $U(R)$. In general, that $P_\gamma$ satisfy
(\ref{brack}) allows one to look for a nilpotent BRST operator of
the form \be \label{genq} \QQ=c^\alpha P_\alpha - \half \sum_{n>0}
\phi_{\alpha_1 \dots \alpha_n\alpha_{n+1}}^{\beta_1 \dots
\beta_{n}}(R)\,
 c^{\alpha_1} \dots c^{\alpha_n} c^{\alpha_{n+1}}\,b_{\beta_1} \dots b_{\beta_{n}}\,,
\ee where ghosts $c^\alpha$ and $b_\alpha$ obey
(\ref{anticomrel})\,, $\phi_{\alpha\gb}^{\gamma}$ are the
``structure functions" of (\ref{brack}) and   $\phi_{\alpha_1 \dots
\alpha_n\alpha_{n+1}}^{\beta_1  \dots \beta_{n}}(R)$ for $n>1$ are
higher structure functions.

  The nonstandard  BRST\, operator $\QQ_r$,
  i.e. the nilpotent operator
\be
 \QQ_{r}^2=0
\ee
  of the form (\ref{genq})
  with some nonzero higher structure functions, constructed
in \cite{gelbrst} from $\mathfrak{sph}_r$ generators $R_\alpha$, is
  \bee\label{fullbrstr}\nn
 \ls\QQ_{r}&=&
   c^\ga{}_\gb P^\gb{}_\ga   +
c_{\ga\gm} P^{\ga\gm} + c^{\ga\gb}P_{\ga\gb}+\sum_{j=1}^r \big( c_j
P{}_{j} + c{}_{j}{}_\ga P{}_{j}{}^\ga\big)\,\,\,\,\\ \nn &&
+c^\ga{}_\gb c^\gb{}_{C} b^{C}_\ga -2c^\ga{}_\gb
c_{\ga{C}}b^{\gb{C}}
+2 c^\ga{}_\gb c^{\gb{C}}b_{\ga{C}}
-4 c^{\ga\gb} c_{\gb{C}} b^{C}{}_\ga\,\,\,\,\\ \nn &&
 - \sum_{j=1}^r
 c_\ga{}^\gb c{}_{j}{}_\gb b{}_{j}{}^\ga
 + \sum_{j=1}^r\! \hhh_j\Big(\!2 c^{\ga\gb} c_{\ga\gb}b_j \!
+\!2 c^{\ga\gb}c_j{}_\gb b_j R_j{}_\ga + 4   c^{\ga \gb} c_{\ga C}
b_j{}^C R_j{}_\gb \,\,\,\,\\ \nn&& - 4  c^{\ga\gb} c_{\gb C}
c_j{}_\ga b_j   b_j{}^{C} - 4  c^{\ga B} c_{\ga C} c_{B{E}}  b_j{}^C
b_j{}^{E}\Big)\,,\,\,\, \eee where \bee\nn  \label{genBr}
P_\alpha:\quad P^{A}{}_{B}&=& R^{A}{}_{B}+  r\half \gd^{A}{}_{B}
,\quad P^{\gm{B}}= R^{\gm{B}} , \quad  P_i^{A} = R_i^{A} ,\quad \nn
P_i{}= R_i{}-\hhhh_i{}\,, \\ \quad P_{\ga\gb}&=&   R_{\ga\gb} -
\sum_{i=1}^r \nu_i{}^{-1} R_i{}{}_\ga R_i{}{}_\gb\,, \eee
  form a ``closed algebra" $\mathfrak{P}_{r}$ with nonzero commutation relations
  that follow from (\ref{comrelsp1}) and (\ref{crsph1})
\bee \nn 
[P^\ga{}_\gb \,,P^{C}{}_{E}] &=& \delta_\gb^{C} P^\ga{}_{E}
-\delta_{E}^\ga P^{C}{}_\gb \,,\\ \nn [P^\ga{}_\gb \,,P^{{C}{E}}]&=&
\delta_\gb^{C} P^{\ga{E}} + \delta_\gb^{E} P^{\ga{C}}\,,
\\ \nn
[P_{\ga\gb},P^{\gm\gn}]&=&\gd_\ga^\gm P_\gb{}^\gn+\gd_\gb^\gm
P_\ga{}^\gn +\gd_\ga^\gn  P_\gb{}^\gm+\gd_\gb^\gn P_\ga{}^\gm +
\sum_j\hhh_j P_j{}                                     \nn
(\gd_\ga^\gm \gd_\gb{}^\gn+\gd_\gb^\gm \gd_\ga{}^\gn )\,\\ \nn
&-&\sum_j\hhh_j(\gd_\ga^\gm R_j{}{}_\gb P_j{}{}{}^\gn+\gd_\gb^\gm
R_j{}{}_\ga P_j{}{}{}^\gn +\gd_\ga^\gn  R_j{}{}_\gb
P_j{}{}{}^\gm+\gd_\gb^\gn R_j{}{}_\ga P_j{}{}{}^\gm ) \,,\\\nn
[P_{\ga\gb} ,P^D{}_{C}]&=& \gd_\ga^D P_{\gb{C}}+\gd_\ga^D
P_{\ga{C}}\,,\\ \nn 
[P_{\ga\gb},P_j^{C }]&=&(\gd_\ga^{C} R_j{}_\gb+\gd_\gb^{C}
R_j{}_\ga)P_j \,,
\\\nn 
[P^{\ga}_{\gb} ,P_j^D ]&=&  \gd_\gb^D P_j^{\ga}\,. \eee

  Specifically, in the case of $r=2$ with
  $\hhhh_1=-\hhhh_2=\hhhh$  the 
  generators $P_\alpha$ of $\mathfrak{P}_{2}$ (\ref{genBr}) are
\bee\nn  \label{genBpm} P_\alpha:\quad P^{A}{}_{B}&=& R^{A}{}_{B}+
\gd^{A}{}_{B} ,\quad P^{\gm{B}}= R^{\gm{B}} , \quad  P_\pm^{A} =
R_\pm^{A} ,\quad \nn P_+{}= R_+{} , \quad  P_-{}= R_-{}-\hhhh {}, \\
\quad P_{\ga\gb}&=&   R_{\ga\gb} - 4 \nu {}^{-1} R_+{}{}_\ga
R_-{}{}_\gb\,. \eee

\section{Standard oscillator realization of the\\ nonstandard  BRST operator}
\label{Oscillator} The appearance of the nonstandard BRST operator
in \cite{gelbrst} was a kind of mysterious and looked nontrivial.
Here we show that it admits a very simple, although nonpolynomial,
equivalent form resulting from the oscillator realization of
$sp(2M)$.

Let \bee\label{genBD} \II_{\ga}{}_{\gb}=  R_{\ga }R_{ \gb}\,,\quad
\II_{\gp}{}^{\gs}\equiv\II^{\gs}{}_{\gp}=\half R_{\gp }R^{ \gs}
+\half   R^{\gs }R_{ \gp}\,,\quad \II^{\gm}{}^{\gn}=R^{\gm}R{}^{\gn}
\,,\quad \II=R\,, \eee where $R_\ga$ and $R$ are  vector fields of
the Heisenberg algebra $\mathfrak{ h}\subset\mathfrak{sph}$, that
satisfy \be \label{R}
  [R_B\,,\,R^A]= \gd_B^A R\,.
  \ee
Denoting the indices of $\mathfrak{sp}$ by single calligraphic
letters $\A\,, \B\,, \dots$, we have
 \bee\label{addcomrel}
[\II_\A{} ,\II_\C{} ]=f{}_{\A  \,}{}_{\C  \,}{}^{\D}\,R\, \II_{\D }
\,,
  \eee
with the structure coefficients $f_{\A  \,}{}_{\C  \,}^{\D}$ to be
read off Eqs.~(\ref{comrelsp1}), (\ref{crsph1}). This is nothing but
the standard oscillator realization of $sp(2M)$ \cite{BG} provided
that the central element $R$ takes some fixed nonzero value, which
is indeed true for the $Q_r$ closed elements as follows from
Eq.~(\ref{fullbrstr}) at $\nu_i\neq 0$.

Since $R$ is nonzero, it is  eligible to introduce operators
$T_\A=R^{-1}\II_{\A}:$ \bee\label{genBDTTT} T_{\ga}{}_{\gb}=
R^{-1}R_{\ga }R_{ \gb}\,,\quad T_{\gp}{}^{\gs}\equiv
T^{\gs}{}_{\gp}=R^{-1}\half\left( R_{\gp }R^{ \gs} +    R^{\gs }R_{
\gp}\right)\,,\quad T^{\gm}{}^{\gn}=R^{-1}R^{\gm}R{}^{\gn} \,,
 \eee
that have the following commutation relations among themselves and
with the vector fields of $Sp(2M)$ \bee\label{addcomrelt} [T_\A{}
,T_\C{} ]=f{}_{\A  \,}{}_{\C \,}{}^{\D}\, T_{\D }\,,
  \eee
  \bee\label{addcomreljt} [R_\A{} ,T_\C{} ]=f{}_{\A  \,}{}_{\C
\,}{}^{\D}\, T_{\D }\,.
  \eee
{}As a result, the operators \be\label{SpO} \Oo_\A = R_\A - T_\A \ee
also fulfill the commutation relations of $\mathfrak{sp}(2M)$
\bee\label{addcomrelOo} [\Oo_\A{} ,\Oo_\C{} ]=f{}_{\A  \,}{}_{\C
\,}{}^{\D}\, \Oo_{\D }
  \eee
and commute to all vector fields of the Heisenberg group \be\nn
[\Oo_\A \,, R_B ]=0\q[\Oo_\A \,, R^B ]=0\q[\Oo_\A \,, R  ]=0. \ee
{}From here it follows nilpotency of  any $Q$ of the form \be
\label{Qnew} Q = Q_\Oo+Q_H\q Q^2=0\,, \ee where $Q_\Oo$ has  the
$\mathfrak{sp}(2M)$ canonical BRST form for the operators $\Oo_\A$
\bee\label{QnewOo} Q_\Oo &=& c^{\ga\gb}\Oo_{\ga\gb} +c^\ga{}_\gb
\Oo^\gb{}_\ga   + c_{\ga\gb} \Oo^{\ga\gb}+
    \\ \nn &&
c^\ga{}_\gb c^\gb{}_{C} b^{C}_\ga -4 c^{\ga\gb} c_{\gb{C}}
b^{C}{}_\ga +2 c^\ga{}_\gb c^{\gb{C}}b_{\ga{C}}   -2c^\ga{}_\gb
c_{\ga{C}}b^{\gb{C}} \phantom{\half} \eee and $Q_H$ is any BRST
operator built from the Heisenberg vector fields.
 In the
case of interest we set \be\label{QHu} Q_H=   c{}_{A} R{}^{A}   +c
(R-\nu). \ee

It turns out that the nonstandard BRST operator of \cite{gelbrst} is
related to $Q $ (\ref{Qnew}) via a canonical change of variables
that preserves the commutation relations between the ghost variables
and vector fields. The new realization of the nonstandard BRST
operator via the oscillator realization of $\mathfrak{sp}(2M)$ not
only fully explains its origin, but also simplifies the relation of
the BRST form of the dynamical equations with its unfolded
formulation. Indeed, the unfolded equations, that result from the
construction of Section \ref{Cartan forms}, applied to the BRST
operator (\ref{QnewOo}),
 are
\bee\label{unfolold}
         \! Df=\!\left(\!d- R^{-1} \go^{\ga\gb}R_{\ga}R_{\gb}
-R^{-1}\half \go^\ga{}_\gb \big( R^\gb{}R_\ga\!+\!R^\gb R_\ga\big) -
R^{-1} \go_{\ga\gb} R^{\ga}R^{\gb}\!\right)\! f\!=\!0. \eee At
$R\neq 0$, these are just the $Sp(2M)$ invariant unfolded equations
proposed in \cite{BHS} where the $\go$ dependent terms were
interpreted as the $Sp(2M)$ connection in the Fock module.

A somewhat unusual feature of the new operator $Q$ is its
nonpolynomiality in $R$, that was not allowed in the analysis of
\cite{gelbrst}. Hence the explicit form of the relation between the
two BRST operators is rather involved and also nonpolynomial.

\section{Closed forms from BRST cohomology}
\label{BRST clo}

\subsection{General case}
\label{genc} Here we consider a coordinate independent realization
of the closed forms that underly the construction of HS currents of
\cite{cur} and \cite{gelcur}, using the correspondence between
unfolded and BRST formulations discussed in Section \ref{Cartan
forms}.

Let $Q_{B}$ and $Q$ be two nilpotent operators \be Q_B^2=0\q
Q^2=0\,, \ee where $Q_B$ is a canonical BRST operator associated to
some group $B\subset G$ while $Q$ is not necessarily canonical.

Consider an element \be F= \Omega f\,, \ee where $\Omega$ belongs to
the algebra $A$ generated by $R,c$ and $b$, to which $Q_{B}$ and $Q$
belong, and  has a nonnegative ghost number $p$, while $f$ belongs
to a left $A$-module and satisfies the conditions \be \label{eqf} Q
f=0 \ee and \be \label{f0} b_\alpha f=0\,. \ee Eq.~(\ref{eqf}) is
the dynamical equation obeyed by $f$ while Eq.~(\ref{f0}) implies
that $f$ is $c$-independent and hence should be interpreted as a
zero-form in the unfolded formulation. {}From the equations
(\ref{eqf}) and (\ref{f0}) it follows that \be \label{Pf0}
 P_\alpha f=0\q P_\alpha= \{Q\,, b_\alpha\}\,.
\ee These are the independent equations encoded by the equation
(\ref{eqf}).

 We will
refer to $\Omega$ and $f$ as a $p$-form and $0$-form, respectively,
since they become those in the unfolded interpretation of the model.
In addition it is required that, by virtue of Eqs.~(\ref{eqf}) and
(\ref{f0}), \be \label{QB0} Q_B F \Big |_\N=0 \,, \ee for some
submanifold $\N$of $G $. In accordance with the general analysis of Section
\ref{Cartan forms}, this implies that, for any orbit $O_B$ of $B$
in $G$, the pullback of the $p$-form
 $F= \Omega f$ to   $ \N\bigcap O_B$  is closed
provided that the zero-form $f$ satisfies its field equations. This
acquires the interpretation of the current conservation once  $f$ is
expressed via bilinears of some other fields $C$ as in the  examples
of Section \ref{Bilinear}.

Now, let us discuss the freedom in the definition of $F$. First of
all, to describe a nontrivial charge conservation, $F$ should belong
to $Q_B$ cohomology on $N$. Indeed, from the analysis of
Section \ref{Cartan forms} it follows that
 $Q_B$-exact $F$ leads to an exact form on $ \N\bigcap O_B$,
hence not contributing to the integrated charge.

Another ambiguity originates from \be F_l(\eta_{\,l})= \eta_{\,l}
\Omega f\,,
 \ee
 where $\eta_{\,l}$ is a $Q_B$ closed element of
ghost number zero \be \label{Qle} [Q_B \,,\eta_{\,l}]=0\,.
 \ee
 Clearly, $F_l(\eta_{\,l})$
satisfies all conditions on $F$. Note, however, that not every $Q_B$
closed $\eta_{\,l}$ leads to a nontrivial result because some
contributions may vanish by virtue of the equations (\ref{eqf}) and
(\ref{f0}).

Alternatively, a $Q$ closed $\eta_r$ of ghost number zero makes it
possible to define \be F_r(\eta_r)=  \Omega \eta_r f\,, \ee where
\be\label{Qet} [Q\,,\eta_r ]=0\,. \ee The meaning of $\eta_r$ is
simple. Various $\eta_r$ describe genuine symmetries of the equation
$Qf=0$ just mapping one solution to another. Their interpretation is
less trivial in terms of the rank one fields $C_1$ and $C_2$ used to
compose a rank two field $f\sim C_1 C_2$ that leads to nontrivial
charge conservation in the rank one model. In this case the
insertion of $\eta$ affects essentially the form of the bilinear
current, leading to different conserved charges. From this point of
view, $\eta$ describe symmetries of the rank one field equations
induced by the conserved currents upon quantization. More precisely,
symmetries of rank one fields are described by the $Q_B$ cohomology
of the space of $F_r(\eta_r)\cup F_l(\eta_{\,l})$. As shown in the
next section, in the cases of interest the ambiguities due to
$\eta_{\,l}$ and $\eta_r$ are equivalent, \ie $F_r(\eta_r)=
F_l(\eta_{\,l})$.

Another benefit of introducing parameters $\eta$ into the definition
of conserved charges is that they allow us to extend them to a
larger space. Suppose for simplicity that (\ref{QB0}) is true for
any $\N$, \ie $\N=G$. The equation (\ref{QB0}) then implies that
$d_B F=0$ on $G$. This allows us to integrate $F$ over
  submanifolds
of   any orbit of $B$ in $G$   (e.g., of  $B$
itself).

Let us introduce an   operator $\Pi_B$ that solves the equation \bee
\label{MtokSph20}
  {Q}_{G}\,  \Pi_B\,=  \Pi_B \,{Q}_{B}\,.
  \eee
For any  $\eta_{\,l}$, the form
  \be
  \Phi =  \eta_G \Omega\, \,f\q \eta_G = \Pi_B \eta_{\,l}
  \ee
is ${Q}_{G}$ closed since \be \label{MtokSph200} Q_G\eta_G=\eta_G
Q_B \,:\qquad
 {Q}_{G} \eta_G \Omega\, \,f=0\,.
\ee

It should be stressed that it is not a priori guaranteed that the
equation (\ref{MtokSph20}) admits a global solution on $G$. This
construction is useful to relate conserved charges that may result
from integration over close surfaces in $G$ which usually give
equivalent results modulo redefinition of the symmetry parameters
$\eta$. In fact, the relation between different parameters is just
governed by the equation (\ref{MtokSph20}) that leads to different
restrictions of $\eta_G$ to different surfaces in $G$. Also let us
stress that, contrary to the equation (\ref{Qle}), the first of the
equations (\ref{MtokSph200}) is not solved by $\eta_G=const$, \ie
the equation (\ref{MtokSph200}) reconstructs appropriate dependence
of $\eta_G$ along the directions transversal to orbits of $B$. Also
note that, for different  subgroups $B$, this procedure may lead to
different results related by a redefinition of $\eta$.

 \subsection{Nontrivial symmetries}
\label{Parameters}
\renewcommand{\A}{\alpha}
\renewcommand{\B}{\beta}
\renewcommand{\C}{\gamma}
\renewcommand{\D}{\mu}
\renewcommand{\E}{\nu}
\renewcommand{\F}{\kappa}

The  conserved currents of \cite{cur,gelcur}
 bilinear in HS fields in the generalized matrix space-time
$\M_M$ depend on constant parameters
  $\eta^{ B_1\ldots B_n}{}_{ A_{ 1} \ldots  A_{m}}\,$
 associated to different HS
symmetry parameters. Let us show how these parameters result from
the general construction of  the previous subsection.

First of all we observe that in our construction $Q_B$, $Q$ and
$\Omega$ are built from  the right $SpH_2$ vector fields $R_\A$.
Hence, any parameter $\eta \in U_2({\XR_\mu})$ composed of the left
$SpH_2$ vector fields ${\XR_\mu}$ obey the properties of both
$\eta_{\,l}$ and $\eta_r$ which, in turn, should be identified
within this class because $\XR_\mu$  commute to $\Omega$.

However, the space of effective symmetry parameters is smaller than
$U_2({\XR_\mu})$ because some of $\eta \in U_2({\XR_\mu})$ act
trivially on  $f$ that satisfy the equations (\ref{Pf0}). In other
words, some elements of $V\in U_2({\XR_\mu})$ can be represented in
the form \be V=\sum_\mu a_\mu(x, \partial) P_\mu \ee where $P_\mu
\in \mathfrak{P}_2$ (\ref{genBpm}) and $a(x,\partial)$ are some
differential operators on  $G$. Since $\Phi_i(P)$ commute to
$\XR_\mu$, the space $I$ of $V$ forms a two-sided ideal of
$U_2({\XR_\mu})$. The quotient algebra $S= U_2({\XR_\mu})/I$
describes true symmetries of the space of solutions of the equations
(\ref{Pf0}).

\renewcommand{\A}{\mathcal{A}}
\renewcommand{\B}{\mathcal{B}}
\renewcommand{\C}{\mathcal{C}}
\renewcommand{\D}{\mathcal{D}}
\renewcommand{\E}{\mathcal{E}}
\renewcommand{\F}{\mathcal{F}}

Let  $S_\mu{}^\nu (x)$  relate the right vector fields $R_{\nu}$ of
a Lie algebra   $\mathfrak{g}$  of some Lie group $G$ to the left
ones $\XR_{\mu}$\,, \bee\label{SLIE} \XR_\mu=S_\mu{}^\nu(x) R_\nu.
\nn\eee Clearly, in any coordinates $x^\kappa$ on $G$,
$$S_\alpha{}^\beta(x)=
\XR_\alpha{}^\kappa (x)\,R^{-1}{}^\beta{}_\kappa(x)\q
\mbox{where}\quad
 \XR_\beta =\XR_\beta{}^\kappa\f{\p}{\p x^\kappa}\q
 R_\beta = R_\beta{}^\kappa\f{\p}{\p x^\kappa}\,.
$$
{}From the Lie algebra commutation relations and mutual
commutativity of left a right vector fields  it follows that
 \bee\label{RSG}\ls
[R_\mu\,, S_{\alpha } {}^\beta ]
 =- f_\mu\,{}_{\lambda}{}^\beta S_{\alpha } {}^\lambda \,,\quad
 [\XR{}_{\gamma}\,, S_{\beta} {}^\mu ]
=f_{\gamma\,\,}{}_{\beta\,\,}{}^\nu S_\nu{}^\mu
 \,,\quad
   -f{}_\beta{}_\mu{}^\nu \,S_{\alpha } {}^\beta S_{\gamma } {}^\mu  =
 f_\alpha{}_\gamma{}^\beta S_{\beta } {}^\nu\,, \,\,\,etc.
\eee

It is convenient to use
 the short-hand notation  $$R_a=( R_A\,,\,R^A)\,,\quad
 R_{ab}=( R_A{}_B\,,\,R^A{}_B \,,\,R^A{}^B)\,,\quad
 L_a= (L_A\,,\,L^A)\,,\quad etc.
$$
Let us now consider the case of $SpH$ starting with the relations
between the vector fields $R_a,\,\,R$(\ref{rightgzr}) and
$\XR_a,\,\,\XR$(\ref{leftgzr}) listed in Section \ref{Vector fields}
in the particular coordinates  (\ref{coorsp}), (\ref{coorHsr}) for
the case of $r=1$.

One can see that \be\label{SHeis}\XR_a=S_a{}^b R_b + S_a{}  R\q \XR=
R\,,\ee
 where
 \bee
 S_a{}^b= \label{matrixSH}
\beee{ r }   \left( \beee{ c c }
 -\D&-\C\D         \phantom{\half}\\
X\D&(\A+\C\D X  )    \phantom{\half}\\
 \eeee \right)
 \eeee\q
 S_a{}= \beee{ r }   \left( \beee{ c      }
 -  w \phantom{\half}\\
  y \phantom{\half} \\
  \eeee \right)
 \eeee
\,. {} \eee It is convenient to use Eqs.(\ref{RSG}) along with
(\ref{crsph1}) to obtain
 \bee\label{RSH}
 [R_{ab\,}\,, S_{c } {}^d  ]
 =-S_{c } {}^e f_{ab\,\,}\,{}_{e}{\,}^d  \,,\quad
[R_{ab}\,, S_{c } {} ]=0\,,\quad[R_m\,, S_{a } {}^b ]=0\q \\ \nn
[R_m\,, S_{a } {}  ] =-S_{a } {}^n f_m\,{}_{n\,}{}^\cdot  \,,\quad
  - f{}_b{\,\,}_m{\,\,}^\cdot \,S_{a } {}^b S_{c} {}^m  =
 f_a{}_c{}^\cdot S  {} \,,\qquad
\eee where $f_a{}_c{}^\cdot$ is defined via $[R_a\,,R_b]=
f_a{}_c{}^\cdot R$.
Taking into account (\ref{RSH}) and antisymmetry of $f_a{}_c{}^\cdot$
in $a$ and $c$, from (\ref{SHeis}) it is elementary to obtain
\bee\label{sphlp}  \XR_{(a}\XR_{c)}=   S_a{}^b S_c{}^d R_{(b} R_{d)}
+ S_{(c}{}^d S_{a)}{} R_d R + S_a{}S_c{}R R\,. \eee Denoting \bee
\label{SSS} \beee{ r c l r c l r c l    } S_a{}_c{}^b{}^d
&=&S_{(a}{}^b S_{c)}{}^d,\qquad& S_a{}_c{}^d&=&S_{(c}{}^d S_{a)} ,&
S_a{}_c{}&=&S_a{}S_c{},\\ \nn \XRp{}_{ab}&=&\XR^{-1}\XR_{(a}\XR_{b)}
,&
  \XRp{}_{a}&=& \XR_{a}  ,&
 \XRp&=&\XR\,,\quad\\
 \Xp_{ab}&=&R^{-1}R_{(a}R_{b)},&
  \Xp{}_{a}&=& R_{a}  ,&
 \Xp&=&R\,\quad
\eeee \eee and using that $L=R$ by virtue of (\ref{SHeis}), we
obtain that for all $\mathfrak{sph}$ indices $\alpha$, $\beta$
\bee\label{Salphabeta}\XRp{}_{\alpha}   = S_{\alpha} {}^\beta
\Xp_\beta\,. \eee
 Note that, as mentioned in Section \ref{Oscillator},
 $\Xp_{\alpha}\equiv T_{\alpha}$ (\ref{genBDTTT})
form  $\mathfrak{sph}$, as well as $\XRp{}_{\alpha}$. Moreover,
analogously to (\ref{addcomrelt}), (\ref{addcomreljt})
\bee\label{cocococ}
[\Xp_{\alpha},\Xp_\beta]&=&[R_{\alpha},\Xp_\beta]=f_{\alpha\,\,}{}_{\beta\,\,}{}^{\gamma}
\Xp_\gamma\q\\ \nn
[\XRp_{\alpha},\XRp{}_\beta]&=&[\XR_{\alpha},\XRp{}_\beta]=f_{\alpha\,\,}{}_{\beta\,\,}{}^{\gamma}
\XRp{}_\gamma\q\\ \nn
[\XRp{}_{\alpha},\Xp_\beta]&=&[\XR_{\alpha},\Xp_\beta]=[\XRp{}_{\alpha},R_\beta]=0\,.
\eee As a result, it follows that $S_{\alpha} {}^\beta$
(\ref{SHeis}), (\ref{SSS}) satisfy (\ref{RSG}) where
$f_\alpha{}_\gamma{}^\beta$ are $\mathfrak{sph}$ structure
constants. Indeed,
 from (\ref{Salphabeta}) along with (\ref{cocococ}) it follows for example that
\bee\nn
[R_{\gamma},\XRp{}_{\alpha}]&=&   
( [R_{\gamma}, S_{\alpha} {}^\mu ]+S_{\alpha} {}^\beta
f_{\gamma\,\,}{}_{\beta\,\,}{}^\mu)\Xp_\mu=0 \q
\\ \nn
[\XR{}_{\gamma}\,, \XRp_{\beta}   ]&=&
f_{\gamma\,\,}{}_{\beta\,\,}{}^\nu S_\nu{}^\mu\Xp_\mu
  =[\XR{}_{\gamma}\,, S_{\beta} {}^\mu ]\Xp_\mu\,.\qquad
\eee

Therefore the vector fields
 \bee\nn\widetilde{\XR}{}_{\alpha} =
S_{\alpha }{}^\beta  R_\beta\, \eee satisfy $\mathfrak{sph}$
commutation relations and  commute to $R_\mu$.  Hence,
 $\widetilde{\XR}{}_{\alpha}$ form left vector fields on $SpH.$
Recall that, by construction,
 $\widetilde{\XR}{}_{a}= {\XR}{}_{a} $ and $\widetilde{\XR}{} = {\XR}{}.$

As a result, from (\ref{genBDTTT}), (\ref{SSS}) and (\ref{sphlp}) it
follows that \bee\label{S R left}  \Oo^{(l)}{}_{a }{}_{b }
  = S_{a }{}_{b }{}^e{}^{d }\Oo_e{}_d\q\mbox{where \,\,}
  \Oo^{(l)}{}_{a b }=\widetilde{\XR}{}_{a }{}_{b }
-\XRp{}_{a b}\q \Oo_{a }{}_{b } =R{}_{a }{}_{b } - \Xp{}_{a b}\,.
\eee Since $\Oo_{a }{}_{b }$ annihilates solutions of (\ref{Pf0}),
 $\Oo^{(l)}_{a b }\in I$. Hence the symmetry algebra $S$ is
generated by  $\XR{}_a$. This result  extends to any rank $r$.

\subsection{$M-$ forms}
\label{cloM} Consider the Lie algebra $\mathfrak{sph}_2$
 and associated ghosts (\ref{antisph}). Let
 \bee \ls
\Omega\!\!&=&\!\!\! \f{1}{  M !}\,\gvep_{\ga_1 \dots  \ga_M}
(c{}^{\gb_1}{}^{\ga_1} R{}_-{}_{\gb_1}+\chalf c{}{}_+{}^{\ga_1}
R{}_-{} )\!\dots\! (c{}^{\gb_M}{}^{\ga_M} R{}_-{}_{\gb_M}+\chalf
c{}{}_+{}^{\ga_M} R{}_-{} )
\equiv\label{c-tok}\\\nn \\
 \label{goodc-tok}
\!\!&\equiv&\!\!\!\!\sum_{k=0}^M   \! \f{ 4^{k- M }}{ k!(M-k)!}
\gvep_{A_1\dots A_M} c{}^{B_1}{}^{A_1}\!\!\dots\!
c{}^{B_k}{}^{A_k}\, c_+{}^{A_{k+1}}\!\!\dots\!
c_+{}^{A_M}
 R{}_-{}_{B_1}\!\dots\! R{}_-{}_{B_k} (R_-)^{M-k},
  \eee
     where
 $\gvep_{B_1 \dots  B_M}$ is the totally antisymmetric Levi-Civita
symbol and
 $"\pm"$ variables of $\mathfrak{sph}(2M)_2$ are defined in
(\ref{sphpm}) and (\ref{sphpmcb}).

The minimal subgroup $B$ of $SpH_2$ that allows to consider the form
(\ref{c-tok}) is $\NNXY$, where $\TT$ is the Abelian subgroup of
translations  (\ref{4}) while $\HHH(y_+) $
  is the subgroup of $H_{M}\times H_{M}$
with the coordinates $y_+{}^A$, introduced in (\ref{sphpm}). 
The respective BRST operator is \bee\label{NNXY}
Q_B&=&c^{\ga\gb}R_{\ga\gb}
  +c_+{}^\ga{}  R_+{}_\ga \equiv\\ \nn&& c^{\ga\gb}P_{\ga\gb} -\hhh c_+{}^{A} R_+{}_{A}P_-
+ 4 \hhh\big(
 c^{\ga\gb}   R_-{}_\gb
 +\chalf c_+{}^\ga R_-\big)R_+{}_\ga\,,
 \eee
where $P_\alpha$ are given in (\ref{genBpm}).

Let from now on $f$ be a ghost independent and $\QQ_2$ closed
function \be\label{Q2clof} \QQ_2 f=0\,, \ee  where $\QQ_2$ is the
nonstandard   BRST operator (\ref{fullbrstr}). Since
$[Q_B\,,\,\Omega]=0$,  $ \Omega  f $   turns out to be
 ${{Q}_B }$ closed as a consequence of (\ref{Pf0}) and the fact
 that $(c^{\ga\gb}   R_-{}_\gb +\chalf c_+{}^\ga R_-)^{M+1}=0$.

Let any  $B\subset SpH_2$ such that
$$Q_B \Omega f=0 $$
be called closure subgroup. The maximal closure subgroup turns out
to be  $SpH_+=SpH\htimes \HHH(y_+\,, u_+\,, w_+  )$, where $\HHH  $
has coordinates $(y_+\,, u_+\,, w_+  )$ (\ref{sphpm}). The
corresponding  BRST operator is
 \bee\label{QSpH+}  {{Q}_{SpH_+}} =
 c^{\ga\gb}R_{\ga\gb}
+c^\ga{}_\gb R^\gb{}_\ga   + c_{\ga\gb} R^{\ga\gb} +c_+ R_+{}     +
c_+{}_{A} R_+{}^{A} + c_+{}^{A} R_+{}_{A}
\\ \nn \phantom{\half}
+c^\ga{}_\gb c^\gb{}_{C} b^{C}_\ga -4 c^{\ga\gb} c_{\gb{C}}
b^{C}{}_\ga
+2 c^\ga{}_\gb c^{\gb{C}}b_{\ga{C}} 
 -2c^\ga{}_\gb c_{\ga{C}}b^{\gb{C}}
\\ \nn-c_\ga{}^\gb c_+{}_\gb b_+{}^\ga\,
+2c_\ga{}_\gb c_+{}^\gb b_+{}^\ga\, +c_\ga{}^\gb c_+{}^\ga b_+{}_\gb
-2c^\ga{}^\gb c_+{}_\ga b_+{}_\gb\, + \half c_+{}_\ga{}  c_+{}^\ga
b_+{}
 \,. \eee
Indeed, using  the  relations and (\ref{genBpm}) we obtain \bee
\label{tokprf10} \Big\{{Q}_{SpH_+}\,,\,
 c{}^{D}{}^{E}
R{}_-{}_{D}+\chalf c{}{}_+{}^{E} R{}_-{} \Big\}
=-    {2 c_{\ga{D}} c{}^{D}{}^{E}
 R{}_-{}^{\ga }}
  +  c_\gb{}^{E}\big( c^{\gb{D}}
R{}_-{}_{D} +\chalf
      c_+{}^\gb  R{}_-{}
\big)\,    \eee and
  \bee \label{tokprf11}
 {{Q}_{SpH_+}}
f= 4\, \,\hhh\big(
 c^{\ga\gb}   R_-{}_\gb
 +\chalf c_+{}^\ga R_-\big)R_+{}_\ga
f-  c^\ga{}_\ga f\,.\eee {}From here it follows that $
{{Q}_{SpH_+}}\Omega  f=0 $. Clearly, any subgroup of $SpH_+$ that
contains the minimal closure subgroup ${\TT\htimes \HHH(y_+)}$, like
e.g. ${P_u\htimes \HHH(y_+)}$ and ${P_u\htimes \HHH(y_+ \,, w_+\,,
u_+\,  )}$, is also a closure subgroup of $\Omega f$. (Recall that
$P_u$ is the upper  parabolic subgroup  of $Sp(2M|\mathbb{R})$
(\ref{p}).)

Using   Eqs.~(\ref{Q=d}), we obtain
  from (\ref{goodc-tok}) that
 \bee \label{dtokcartan}Q_B \Omega f=0 \Rightarrow
   d\big|_{B}\, \, \widetilde{\Omega}\,\,f =0
,\eee where  $d\big|_{B}$ is the exterior differential on $B$, \bee
\label{tokcartan}   \widetilde{\Omega} &=&
  \sum_{k=0}^M     \f{ 4^{-(M-k)}}{
k!(M-k)!}\, \gvep_{A_1\dots A_M}\\ \nn&&
\go{}^{B_1}{}^{A_1}\wedge\dots\wedge \go{}^{B_k}{}^{A_k}\wedge
\go_+{}^{A_{k+1}}\wedge\dots\wedge \go_+{}^{A_M}\,
R{}_-{}_{B_1}\dots R{}_-{}_{B_k}
 (R_-)^{M-k}\,
   \eee
and $\go$ are Cartan forms on $B$. In Section \ref{Mforms2}, we show
that the pullback of the form
  $\widetilde{\Omega} $ to $\NNXY\subset SpH_2$ reproduces the
  conserved current of \cite{gelcur} provided that $f$ is bilinear in
solutions of rank one field equations. The form
$\widetilde{\Omega}f$
 provides the coordinate independent
generalization of the  conserved HS currents of \cite{cur,gelcur}.

\subsection{$3M-$ forms}
\label{BRST clo3M}

The correspondence between conserved HS currents in four dimensional
Minkowski space $\M_4$ and those in the ten dimensional matrix space
$\M_4$ was established in \cite{gelcur}. Since the charge in $\M_4$
contains four integrations versus three in  Minkowski space, the
naive reduction with fourth integrations  over a cyclic spin
variable in $\M_4$ gives zero. To make the cycle noncontractible, a
singularity (flux) should be introduced in the spinning space.  It
was suggested in \cite{gelcur} to use for this aim a generalized
$2M-$form current. This was achieved by introducing additional
spinor variables $W$. The corresponding currents  were of the form
$$(d W)^M (W dX +   dY_+  )^M \eta(W,\,\,W X +Y_+)g(W,\,Y_+|\,X)\,,$$
where the parameters $\eta $ were arbitrary  functions of $\,W X
+Y_+\,$ and $\,\,W\,$ while $g(W,\,Y_+|\,X)$ was related to the
stress tensor via the half Fourier transform that replaced
$\dis{\frac{\partial}{\partial Y_-}}$ by $W$.

This generalization allowed us to consider singular parameters
$\eta$ necessary to reproduce the standard $4d$ currents in
Minkowski space, that was hard to achieve in
 the original $M-$form current \cite{cur}, where parameters were
polynomials of $  \dis{ \f{\p}{\p Y_-}}$ and $  \dis{\big(X\f{\p}{\p
Y_-} - Y_+\big)}$. However the geometric meaning of the
construction, and, in particular, of the half-Fourier transform was
not clear in the setup of \cite{gelcur}. Here we introduce a
geometric $3M$-form current construction that reproduces
 that of \cite{gelcur}. In the new setup, the half-Fourier
 transform results from the integration over additional $M$ coordinates.

Consider
 \bee \label{c-tok3}  \Lambda&=&
    \Big(c_-{}^\ga \,c_+{}_\ga \Big)^M \Omega\,,
  \eee
where   $\Omega\,$ is of the form  (\ref{c-tok}).

The minimal subgroup of $SpH_2$ that supports the form
(\ref{c-tok3}) is \be\label{Blambda} B={{\TT\htimes \HHH(y_+\,, w_+
\,, u_+\,\,,y_-    \,, u_- )}}, \ee where $\TT$ is the Abelian
subgroup of translations  (\ref{4}) while $\HHH(y_+\,, w_+ \,,
u_+\,\,,y_-    \,, u_- )\subset H_{M}\times H_{M}$ has coordinates
  $(y_+\,, w_+ \,, u_+\,\,,y_-    \,, u_- )$ (\ref{sphpm}).

   The respective BRST operator   ${Q_B} $ can be written in the form
  \bee \label{QNNXY3}\!\!
   Q_B\!\!&=&\!\!c^{\ga\gb}P_{\ga\gb}
     -\hhh c_+{}^{A} R_+{}_{A}P_-
+c{}_\gb P^\gb{} +c_+ P_+{} +c_- R_-{}       + c_-{}^{A} R_-{}_{A}+
   \\ \nn&&
 4 \hhh\big(
 c^{\ga\gb}   R_-{}_\gb
 +\chalf c_+{}^\ga R_-\big)R_+{}_\ga
  \,\, 
   -2c^\ga{}^\gb c_+{}_\ga b_+{}_\gb\,
+ \half c_+{}_\ga{}  c_+{}^\ga b_+{}\,
  + \half c_+{}_\ga{}c_-{}^\ga b_-{}
,\eee where $P_\mu$ are given in (\ref{genBpm}).

One can easily see that $\left[{Q_B} \,,\, \big(c_-{}^\ga
\,c_+{}_\ga \big)^M \right]=0$ and \bee \label{3MtokNNN}\ls
 {Q_B}  \Lambda  f= \nu c_- \Lambda  f\,
  \eee
provided that ghost a independent $f$ satisfies (\ref{Pf0})\,.

Note that the property  (\ref{3MtokNNN}) holds for a larger group
$B_u ={{P_u\htimes \HHH(y_+\,, w_+ \,, u_+\,\,,y_- \,, u_- )}}$,
where $P_u$ is the upper   parabolic subgroup (\ref{p}). The group
$B_u$ is maximal in the sense that further extensions lead to
additional terms on the  r.h.s of (\ref{3MtokNNN}).

Again, using   Eqs.~(\ref{Q=d}) along with (\ref{goodc-tok}), we
obtain that
 \bee \label{dtokcartan3}
 Q_B \Lambda f= \nu c_- \Lambda  f\, \Rightarrow
   d\big|_{B}\, \,\widetilde{\Lambda}\,\,f = \nu c_- \widetilde{\Lambda}  f\,
,\eee where $d\big|_{B}$ is the exterior differential on $B$,
\bee\label{la}\widetilde{\Lambda}f= \Big(\omega_-{}^A
\omega_+{}_A\Big)^M\,\,\widetilde{\Omega}\,f, \eee
 $\widetilde{\Omega}$ is of the form (\ref{tokcartan}) and   $\go$
are Cartan forms  on $B$.

Although the property (\ref{dtokcartan3}) does not imply the closure
of the $3M$-form $ {\Lambda} \,f$ on $B$, it is closed on any
submanifold $\N$ of $B$ such that $\go_-\Big(\omega_-{}^A
\omega_+{}_A\Big)^M\,\Big |_\N=0$, \ie $\N$ is a  kind of {Pfaffian
surface }. This property will be used in  Section \ref{3Mforms2} to
construct conserved currents.

In fact, the formulas (\ref{dtokcartan3}), (\ref{la}) have the
following interpretation. Consider the construction of Section
\ref{cloM} with the $2M$-form parameters $\phi$ that satisfy the
conditions \be \label{cc} c_-^A \phi =0\q \phi c_-^A =0 \q \phi
c_+^A=0\q c_+^A \phi =0\, \ee and \be [Q_B\,,\phi] = -\nu c_-
\phi\,. \ee Then, the $3M$ form \be \label{cp} \Psi=\phi \Omega f
\ee turns out to be $Q_B$ closed
 \be Q_B \Psi=0\,. \ee
 Here the equations (\ref{cc}) imply that $\Psi$ contains a factor of
$\Big(c_-{}^\ga \,c_+{}_\ga \Big)^M$ while the equation (\ref{cp})
determines the dependence of $\phi$ on the central charge coordinate
$u_-$ in such a way that the form $\Psi$ becomes closed. Note that
this construction is to some extent analogous to that described in
the end of Subsection \ref{genc}.

\renewcommand{\A}{\mathcal{A}}
\renewcommand{\B}{\mathcal{B}}
\renewcommand{\C}{\mathcal{C}}
\renewcommand{\D}{\mathcal{D}}
\renewcommand{\E}{\mathcal{E}}
\renewcommand{\F}{\mathcal{F}}

\section{Bilinear currents}
\label{Bilinear} To show that the differential forms
$\widetilde{\Omega}(\eta,f)$ and $\widetilde{\Lambda}(\eta,f)$
introduced in Section \ref{BRST clo} lead to bilinear conserved
currents of \cite{cur,gelcur} we need manifest expressions for the
Cartan forms on $SpH{}_2 $. The straightforward computation  gives
 \bee\label{cartansph2}
\go^{\ga\gb}&=&-\f{1}{2 } \D{}_{ C }{}^{\ga}\D{}_{E }{}^{\gb}d
X^{EC} \,,\\ \nn \go{}_{F}{}^{\gb}&=& -\C_F{}_\ga \D{}_{ C
}{}^{\ga}\D{}_{E }{}^{\gb}d X^{EC} -\D{}_{ F }{}^{\ga}d\A
^{\ga}{}_{\gb}\,,\\ \nn \go{}_{C D}&=& -\half \C_C{}_\gb
\C_D{}_\ga\D{}_{ C }{}^{\ga}\D{}_{E }{}^{\gb}d X^{EC}
-\C_{C\gb}\D{}_{ D }{}^{\ga}d\A ^{\gb}{}_{\ga} +\half d \C_{C D}
\,,\\ \nn \go{}_+{}^{\ga}&=&-\D{}_{ B }{}^{\ga} dy{}_+{}^{ B }+
\D{}_{ B }{}^{\ga}  X^{ B C} d  w{}_+{}_C\,,\\ \nn
\go{}_+{}_{\gb}&=&- \C_\gb{}_\ga\D{}_{ D }{}^{\ga} dy{}_+{}^{ D }
+(\A{}^C{}_\gb+\C_\gb{}_\ga\D{}_{ D }{}^{\ga}  X^{ D C})
 d w{}_+{}_C\,
 \,,
 \eee
 \bee \nn
 \go{}_+{}&=&- \chalf{w{}_+{}_\gb}  dy{}_+{}^\gb
 + \chalf{y{}_+{}^\gb}  d w{}_+{}_\gb
 - \chalf{w{}_-{}_\gb}  dy{}_-{}^\gb
 + \chalf{y{}_-{}^\gb}  d w{}_-{}_\gb -\half du{}_+{}
\,,\\ \nn \go{}_-{}^{\ga}&=&-\D{}_{ B }{}^{\ga} dy{}_-{}^{ B }+
\D{}_{ B }{}^{\ga}  X^{ B C}
 d w{}_-{}_C\,,
 \\ \nn
  \go{}_-{}_{\gb}&=&- \C_\gb{}_\ga\D{}_{ D }{}^{\ga} dy{}_-{}^{ D }
+(\A{}^C{}_\gb+\C_\gb{}_\ga\D{}_{ D }{}^{\ga}  X^{ D C})
 d w{}_-{}_C\,
 ,\\ \nn
 \go{}_-{}&=&- \chalf{w{}_+{}_\gb}  dy{}_-{}^\gb
 + \chalf{y{}_+{}^\gb}  d w{}_-{}_\gb
-\chalf w_{-}{}_{\gm}d  y_{+}{}^{\gm}+ \chalf y_{-}{}^{\gm}d
w_{+}{}_{\gm}
 -\half d u{}_-{}.
\eee
 Let us consider the cases of $M$-forms and $3M$ forms separately.

\subsection{ $M$--forms}

\label{Mforms2}
 To obtain the manifest formula for
$\widetilde{\Omega}(\eta, f)$, note that
 the general solution of rank two equations  (\ref{Pf0})  is
 \bee\label{ungensol}f= \det(\A) \exp
 \Big(\chalf \hhhh \big(2u_- + w_-{}_{B} Y_+{}^{B}+ w_+{}_{B} Y_-{}^{B}
\big)\Big)
    F  (Y_+\,,Y_-|X  ), \eee
where $$ Y_{\pm}{}^{B}=y_{\pm}{}^{B}-
 w_{\pm}{}_{{A}}  X{}^B{}^{A}$$
and $ F(Y_+\,,Y_-|X )$ is any solution of the equation
 \bee \label{CurYY}
 \Big(\f{\p}{\p X^{AB}}
+2 \hhh\,\, \f{\p}{\p Y_{-}{}^{(B}} \,\f{\p}{\p Y_{+}{}^{A)}}
 \Big) F=0\,.
\eee

 Using (\ref{cartansph2})  along with 
(\ref{gz-F}) and (\ref{ungensol}) 
we obtain  from (\ref{tokcartan}) \bee \label{tokcartango2}
\widetilde{\Omega}(\eta,f) &=&
         2^{-  M }
\Big( d X{}^{B }{}^{A }
  \f{\p}{\p y_-{}^{B }}
-\half \nu    \big(d Y{}_+{}^{A}+\half w_+{}_{B  }d X^{AB}\big)
\Big)^M
  \eta\\ \nn&& \exp
 \Big(\chalf \hhhh \big(2u_- \!+ w_-{}_{B} Y_+{}^{B}\!+ w_+{}_{B} Y_-{}^{B}
\big)\Big)
    F    (Y_+\,,Y_-|X  )
    \,,
   \eee
where  $\eta $ is a free parameter of HS symmetries. That the
expression (\ref{tokcartan}) is independent of the coordinates
$\A_A{}^B$ and $ C_{AB} $   is not accidental, being a consequence
of its $Q_{SpH_+}$ closure.

To make contact with the bilinear currents of \cite{cur}, consider
$B=\TT\htimes\HHH(y_+)\subset SpH_2$. Then \bee \label{tokcartanB}
\widetilde{\Omega}(\eta,f)\Big|_{B} =
         2^{-  M }
\Big( d X{}^{B }{}^{A }
  \f{\p}{\p y_-{}^{B }}
\!  -\!\half \nu    d y{}_+{}^{A  } \Big)^M
  \! \!\eta        F   (y_+\,,y_-|X  )\Big|_{y_-=0}
    \,.
   \eee

As  mentioned in Section \ref{Parameters}\,, $\eta$ is a polynomial
in the operators 
   \bee\nn
\XR_{+}{}_{\ga} &=&  -\D_A{}^B R_+{}_B  -\C_{BC}\D_A{}^C
R_+{}^B- \half w_+{}_A R_+{} -\half R_-{} w_-{}_A  ,\quad \\
\XR_{+}{}^{\ga} &=&  + X^{B C}\D_C{}^A  R_+{}_B +\A^A{}_E  R_+{}^E
+\D_C{}^A C_{EB} X^{B C}  R_+{}^E    +\half y_+{}^A  R_+{}
   +\half R_-{} y_-{}^A   ,\nn
\\ \XR_{-}{}_{\ga} &=&
-\D_A{}^B  R_-{}_B -\C_{BC}\D_A{}^C  R_-{}^B -\half w_-{}_A  R_+{}
-\half R_-{} w_+{}_A ,\quad   \label{leftHeis}
\\ \nn \XR_{-}{}^{\ga} &=&
+ X^{B C}\D_C{}^A  R_-{}_B+\A^A{}_E   R_-{}^E +\D_C{}^A C_{EB} X^{B
C}  R_-{}^E +\half y_-{}^A  R_+{} +\half  R_-{} y_+{}^A \,,\eee that
can be obtained using (\ref{matrixSH}).
 Modulo $P {}_\mu$ (\ref{genBpm}),
the pullback to $B$ gives \bee\label{ex3} \XR_{+}{}_{\ga}\big|_B  =
-  R_+{}_A  ,\quad \XR_{+}{}^{\ga}\big|_B =   X^{B A} R_+{}_B
+\half\nu y_-{}^A. \eee

It is easy to see that $\XR_{+}{}_{\ga}$ and $\XR_{+}{}^{\ga}$
(\ref{ex3}) lead to exact forms on $B$. Indeed, since
$\{Q_B\,,b_+{}_A\}=  R_+{}_A$, the $R_+$ dependent part of $\eta$ is
exact. Analogously, setting $\xi^B= X^{AB} b_+{}_A$, we have
$\{Q_B\,,\xi^B\}\big|_B=  X^{AB} R_+{}_A +   c^{AB}b_+{}_{A }$ and
hence \be\label{ex2}  (X^{AB} R_+{}_A +   c^{AB} b_+{}_{A }) \Omega
f\big|_B =Q_B \Upsilon^B\ee for some $\Upsilon^B$. As a result,
effective parameters  depend only on $\XR_{-} $. Using
(\ref{leftHeis}) and neglecting the terms, that belong to the
annihilator $I$ generated by $P{}_\mu$ (\ref{genBpm}), we have
  \bee\label{leftpoP}
\XR_{-}{}_{\ga}\big|_B  \simeq  -  R_-{}_A -\half\nu w_+{}_A  ,\quad
\XR_{-}{}^{\ga}\big|_B  \simeq   X^{B A}  R_-{}_B +\half \nu y_+{}^A
\,. \eee Finally, using
 (\ref{gz-F}) we conclude that nontrivial parameters are
 arbitrary functions of
 \bee\label{param-}
\f{\p}{\p y_{-}{}^{B}} ,\quad     X^{B A}\f{\p}{\p
y_{-}{}^{B}}-\half \nu y_+{}^A
 . \eee

 Setting
$ F=  C^+ \left(\f{y_+{}+y_-{}}{2} \big|X\right) C^-
\left(\f{y_+{}-y_-{}}{2} \big|X\right)
 \,,$ where
 $ C^\pm
(y|X)$  satisfy the rank $1$ unfolded equations
 \be\label{unf1}
\left ( \f{\p}{\p X^{ A B}} \pm \half\,\,\hhh \f{\p^2}{\p y^ A \p y^
B}\right ) C^\pm(y\big|X) =0\,, \ee we obtain from
(\ref{tokcartango2}) \bee \label{bilinM}\ls
\widetilde{\Omega}(\eta,f) \!=\!
         2^{-  M }
\Big( d X{}^{B }{}^{A }
  \f{\p}{\p y_-{}^{B }}
\!  -\!\half\,\nu    d y{}_+{}^{A  } \Big)^M
  \! \!\eta\Big(\f{\p}{\p y_{-}{}^{B}} ,  X^{B A}\f{\p}{\p
y_{-}{}^{B}}\!-\!\half \nu y_+{}^A\Big)
    F (y_\pm|X)  
    ,
   \eee
which, up to a constant, is the bilinear current of \cite{gelcur}.

\subsection{$3M-$ forms}
\label{3Mforms2} In \cite{gelcur} we introduced the generalized
$2M$-closed forms which allowed us to reproduce the usual HS charges
of Minkowski space. However, neither geometric meaning of the $M$
additional coordinates nor the origin of the half-Fourier transform
applied in \cite{gelcur} was not clear in that paper. Here we show
that the $3M$ form (\ref{la}) of Subsection \ref{BRST clo3M}
naturally reproduces the results of \cite{gelcur}.

 Let
$ \N=\TT\htimes \HHH(y_+\,, w_+\,, u_+\,\,,y_-     \,,
u_-)\Big|_{u_-=const}$. Using (\ref{cartansph2})  along with
(\ref{gz-F}) and (\ref{ungensol}), we obtain  from (\ref{la}) \bee
\label{tokcartan3N} \widetilde{\Lambda}(\eta ,\,f)\big|_{{\N }} &=&
2^{-  M }
 \Big( d X{}^{B }{}^{A }
  \f{\p}{\p y_-{}^{B }}
-\half \nu    \big(d Y{}_+{}^{A}+\half w_+{}_{B  }d X^{AB}\big)
\Big)^M
 \times\\ \nn&&\Big(
dy{}_-{}^{\ga }   d w{}_+{}_\ga\Big)^M \eta\, \exp \Big(\f{\hhhh}{4}
w_+{}_{B} y_-{}^{B}    \Big) F (Y_+\,,Y_-|X  )\,. \eee
 Here $F=F (Y_+\,,Y_-|X )$ is any solution of (\ref{CurYY}) and
$ \eta   $ is a function  of  $L_\pm$ (\ref{leftHeis}) up to the
terms that belong to the annihilator $I$ generated by $P{}_\mu$
(\ref{genBpm}).
 Using that
\be \go_- \Big( dy{}_-{}^{\ga }   d w{}_+{}_\ga\Big)^M\,\Big|_\N= 0
\ee by virtue of (\ref{cartansph2}) we obtain that
  \bee \label{dtok3zero}    d \widetilde{\Lambda}f  \big|_{\N } =0\qquad
  \mbox{for any } \QQ_2 \mbox{ closed  }f\,.
 \eee
 Analogously to the case of $M-$forms of Subsection \ref{Mforms2},
one can see that the dependence of the parameters on
$\XR_{+}{}^{\ga} $ and $\XR_{+}{}_{\ga}$ leads  to exact forms.

Consider the
  family  of surfaces    $\N  $ of the form
\be\label{sigmapm}
  \N=\mathbb{R}^M(y_-)\times \mathbb{R}^M(w_+)\times \sigma_{(M)}\,,
\ee where $\sigma_{(M)}$ is any $M$-dimensional surface. In other
words, we consider only such surfaces that their volume forms
contain $\Big( dy{}_-{}^{\ga }   d w{}_+{}_\ga\Big)^M $.

It can be shown that  all terms of the form
$\widetilde{\Lambda}(\eta\,,\,f)\big|_{{\N}  }$, that contain
 $\f{\p}{\p y{}_-{}^{B }}$, are
exact on $\N$.   Indeed, it is evident that
\bee\nn 
\gvep_{A_1\dots A_M}\,\, dy{}{}^{\ga_1} \dots dy{}{}^{\ga_M }
\f{\p}{\p y{}{}^{B }} \Phi   = M\, d\,\,\,\, \Big( \,\, \gvep_{B
\,A_2\dots A_M}\,\, dy{}{}^{\ga_2} \dots dy{}{}^{\ga_M }\Phi
 \,\Big)
\, \eee provided $\ \Phi $ is any   form  on $R^M(y)\times A(x)$ .
Hence, from (\ref{tokcartan3N}) we obtain that
 \bee\label{bezddy}
\ls \widetilde{\Lambda}(\eta\,,\,f)\big|_{{\N}} &\sim&
     \f{(- \nu)^{   M }}{4^M}
 \Big( dy{}_-{}^{\ga }   d w{}_+{}_\ga\Big)^M
\Big(
      \, d Y{}_+{}^{A  }+\half w_+{}_{B  } d X^{AB} \Big)^M\times
\\ \nn &&  \eta \,\exp
 \Big(\half \hhhh  \big(u_- +  \half w_+{}_{B} y_-{}^{B} \big)  \Big)
    F (Y_+\,,y_-|X  )\,,
\eee where $ \eta   $ is a function  of the operators $L_\pm$
(\ref{leftHeis}) modulo $P{}_\mu$ (\ref{genBpm}) and modulo terms
that contain  $\dis{\f{\p}{\p y{}_-{}^{B }}}$. {}From
(\ref{leftpoP})\,, (\ref{gz+F}) and (\ref{gz-F}) it follows that the
'effective' parameters are functions of
   \bee \nn
\XR_{-}{}_{\ga}  \simeq
 -\chalf\nu w_+{}_A  ,\quad
\XR_{-}{}^{E}  \simeq     X^{A E}\chalf\nu w_{+}{}_{\ga}+\half \nu
Y_+{}^A \,. \eee This leads to the expression for the current used
in \cite{gelcur} upon complexification to the Siegel space. The
half-Fourier transform results from the integration over $y_-^A$.

Indeed, setting $$u_-=0\q -i\hbar=\chalf \nu\q F(Y_+\,,y_-|X  )=  C^+
\left(\f{Y_+{}+y_-{}}{2} \big|X\right) C^- \left(\f{Y_+{}-y_-{}}{2}
\big|X\right)
 \,,$$ where
 $ C^\pm
(y|X)$  satisfy the rank one unfolded equations
 (\ref{unf1}), the integration of (\ref{bezddy}) over $y_-$ gives
 \bee\label{bezddyF}
\ls \varpi^{2M} &\sim&
       \Big(   d w{}_+{}_\ga\Big)^M
\Big( d Y{}_+{}^{A  }+\half w_+{}_{B  } d X^{AB} \Big)^M
      \eta\,
      \widetilde{F} (Y_+\,,w_+|X  ),
\eee where $\widetilde{F} (Y_+\,,w_+|X  )\,$ is the half-Fourier
transform of $F (Y_+\,,y_-|X  )$
\bee \nn
 \widetilde{F} (Y_+\,,w_+|X  )= (2\pi)^{-M/2} \int\limits_{\mathbb{R}^M}   
(d\,y_-)^{\,M}\,\,\, \exp\left(-i\,\hbar\,\, w_+{}_{B}
y_-{}^{B}\right)\, F (Y_+\,,y_-|X  )\, \eee and $\eta=\eta( w_+{}_A
,  \half X^{A E}  w_{+}{}_{\ga} + Y_+{}^A)$. Up to a constant, the
equation (\ref{bezddyF}) reproduces the bilinear $2M$-form
current of \cite{gelcur}.

\section {Conclusion}
\label{conc}

The construction of currents presented in this paper not only
explains their coordinate-independent origin but also clarifies
their role in the nonlinear unfolded theory. Indeed, the $M$-- and
$3M$--closed forms of Section \ref{BRST clo} are
analogous to the terms that glue zero-form Weyl modules to the
gauge-field modules via unfolded equations of the form
\be \label{M}
dW +\tilde{\Omega} f =0 \qquad \mbox{or}\qquad  dV+\tilde{\Lambda}
f=0\,, \ee
where $f$ belongs to the Weyl module while $W$ and $V$
should be interpreted, respectively, as a $(M-1)$--  and
$(3M-1)$--form on the respective group $B$. This structure, known
since \cite{Ann}, is typical for the unfolded field equations.

In the context of $Sp(8)$ invariant HS theory analogous
problem was recently considered in \cite{33} where terms of this type were
obtained in the sector of one-forms $W$.
However, the deformation of HS field equations obtained in \cite{33} breaks
the symmetry between dotted and undotted spinors, \ie $GL(4)$
symmetry, and hence essentially differs from the equations (\ref{M})
which provide a new interesting framework for the
deformation of HS field equations. In the case where the rank two
field $f$ is represented by bilinears of the rank one fields $C$ as
in Section \ref{Bilinear}, the equations (\ref{M})
become nonlinear and should
describe the HS current interactions. Given that the realization of
HS currents in $4d$ Minkowski space of \cite{gelcur} was rather
nontrivial, it is tempting to see how the standard HS current
interaction reappears in terms of the twelve-form $\tilde{\Lambda}$.

 \section*{Acknowledgments}
This research was supported in part by
 RFBR Grant No 08-02-00963, LSS No 1615.2008.2
M.V. acknowledges a partial support from the Alexander von Humboldt
Foundation Grant PHYS0167.

\newcounter{appendix}
\setcounter{appendix}{1}
\renewcommand{\theequation}{\Alph{appendix}.\arabic{equation}}
\setcounter{equation}{0}
 \renewcommand{\thesection}{\Alph{appendix}.}
\addtocounter{section}{1}

\medskip

\addtocounter{section}{1}
\addcontentsline{toc}{section}{\,\,\,\,\,\,\,References}
\medskip

\section*{$\rule{0pt}{1pt}$}


\begin{thebibliography}{99}

\baselineskip=18pt
\parindent=0pt
\parskip=-4.5pt

\bibitem{gelbrst} O.~A.~Gelfond and M.~A.~Vasiliev,
 {\it JHEP} {\bf 12} (2009) 21,
      [arXiv:0901.2176  [hep-th]].

\bibitem{BHS}
M.~A.~Vasiliev, \textit{Phys.Rev.} {\bf D66} (2002) 066006, {\tt
[hep-th/0106149]}.


\bibitem{F}C.Fronsdal, \emph{``Massless Particles, Ortosymplectic
Symmetry and Another Type of Kaluza-Klein Theory"}, Preprint
UCLA/85/TEP/10, in Essays on Supersymmetry, Reidel, 1986
(Mathematical Physics Studies, v.8).

\bibitem{BLS1}I.Bandos, J.Lukierski and D.Sorokin, \textit{Phys.Rev.} {\bf D61}
(2000) 045002, {\tt [hep-th/9904109]}.

 \bibitem{Mar}
M.~A.~Vasiliev, \emph{``Relativity, Causality, Locality,
Quantization and Duality in the $Sp(2M)$ Invariant Generalized
Space-Time''}, {\tt [hep-th/0111119]}; in the Marinov's Memorial
Volume, M.Olshanetsky and A.Vainshtein Eds, World Scientific, 2002.

\bibitem{cur} M.A. Vasiliev,
{\it Russ. Phys. J.} {\bf 45} (2002) 670 ({\it Izv. Vuzov, Fizica}
{\bf 45} (2002) N7 23),  {\tt [hep-th/0204167]}.


\bibitem{IB}
I.A.Bandos,
{\it Phys.Lett.} {\bf B558} (2003) 197 {\tt [hep-th/0208110]}.

\bibitem{DV}
V.E.Didenko and M.A.Vasiliev, \textit{J.Math.Phys.} {\bf 45} (2004)
197 {\tt [hep-th/0301054]}.

\bibitem{PST}
M.Plyushchay, D.Sorokin and M.Tsulaia, \textit{JHEP} {\bf 0304}
(2003) 013 [hep-th/0301067];
 \emph{``GL flatness of OSp(1|2n) and higher spin field theory from
 dynamics in tensorial spaces"}, {\tt [hep-th/0310297]}.


\bibitem{tens2}
O.A.Gelfond and M.A.Vasiliev, {\it Theor.Math.Phys.} {\bf 145}
(2005) 35,
[{\tt hep-th/0304020}].


\bibitem{BPST}
I.Bandos, P.Pasti, D.Sorokin and  M.Tonin,
{\it JHEP} 0411:023 (2004) , {\tt [hep-th/0407180]}.

\bibitem{BBAST}
I.Bandos, X.Bekaert, J.A. de Azcarraga, D.Sorokin and M.Tsulaia,
{\it  JHEP} 0505:031 (2005), {\tt [hep-th/0501113]}.

\bibitem{EL} E.Ivanov and  J.Lukierski,
{\it Phys.~Lett.\/} {\bf B624} (2005) 304, {\tt [hep-th/0505216]}.


\bibitem{EI} E.Ivanov,
\emph{``Nonlinear Realization in Tensorial Superspaces and Higher
Spins"}, [{\tt hep-th/0703056}].

\bibitem{33}
M.A. Vasiliev, {\it Nucl.Phys.} {\bf B793} (2008) 469, {\tt
arXiv:0707.1085 [hep-th]}.


\bibitem{gelcur} O.~A.~Gelfond and M.~A.~Vasiliev,
 {\it JHEP} {\bf 03} (2009) 125;
     [arXiv:0801.2191v4 [hep-th]].


\bibitem{Barnich:2004cr}
G.~Barnich, M.~Grigoriev, A.~Semikhatov and I.~Tipunin,
{\it Commun.Math.Phys.} {\bf 260} (2005) 147-181; {\tt
[hep-th/0406192]}.

\bibitem{CLW}
  D.~Cherney, E.~Latini and A.~Waldron,
{\it  ``BRST Detour Quantization,''}
  arXiv:0906.4814 [hep-th].

\bibitem{BGr}
  G.~Barnich and M.~Grigoriev,
{\it   JHEP}  {\bf 0608} (2006) 013
  [arXiv:hep-th/0602166].


\bibitem{Ann}   M.~A.~Vasiliev, {\it Ann. Phys.} (N.Y.) {\bf 190} (1989) 59.

\bibitem{act} M.~A.~Vasiliev,
{\it  Int.\ J.\ Geom.\ Meth.\ Mod.\ Phys.} {\bf 3} (2006) 37
[hep-th/0504090].

\bibitem{solv}
 X.~Bekaert, S.~Cnockaert, C.~Iazeolla and M.~A.~Vasiliev, {\it ``Nonlinear
higher spin theories in various dimensions}, arXiv:hep-th/0503128.


\bibitem{BG} I.~Bars and M.~G\"unaydin, {\it Commun. Math. Phys.}
{\bf 91} (1983) 31.



\end{thebibliography}
\end{document}